\documentclass[a4paper,12pt]{article}

\usepackage{cite}

\usepackage{lscape}
\usepackage{fullpage}

\usepackage{amsfonts}
\usepackage{amsmath}
\usepackage{amssymb}

\numberwithin{equation}{section}

\newcommand{\qandq}{\qquad\mathrm{and}\qquad}
\newcommand{\diag}{\mathrm{diag}}
\newcommand{\HH}{\mathcal{H}}			% for generalized metric in DFT
\newcommand{\MM}{\mathcal{M}}			% for generalized metric in SL5
\newcommand{\EE}{\mathcal{E}}			% for generalized vielbein
\newcommand{\LL}{\mathcal{L}}			% for generalized Lie derivative and Lagrangian
\newcommand{\II}{\mathbb{I}}
\newcommand{\JJ}{\mathbb{J}}
\newcommand{\KKK}{\mathbb{K}}

\newcommand{\dd}{\mathrm{d}}		

\newcommand{\hphi}{\hat{\phi}}			

\newcommand{\bm}{{\bar{m}}}
\newcommand{\bn}{{\bar{n}}}
\newcommand{\bk}{{\bar{k}}}
\newcommand{\bl}{{\bar{l}}}
\newcommand{\bt}{{\bar{t}}}
\newcommand{\bz}{{\bar{z}}}
\newcommand{\ba}{{\bar{a}}}
\newcommand{\bb}{{\bar{b}}}
\newcommand{\bq}{{\bar{q}}}
\newcommand{\bp}{{\bar{p}}}
\newcommand{\bmu}{{\bar{\mu}}}

\newcommand{\tx}{\tilde{x}}
\newcommand{\ttt}{\tilde{t}}
\newcommand{\tz}{\tilde{z}}
\newcommand{\ty}{\tilde{y}}
\newcommand{\tg}{{\tilde{g}}}
\newcommand{\tMM}{\tilde{\MM}}
\newcommand{\tphi}{\tilde{\phi}}
\newcommand{\htphi}{\hat{\tilde{\phi}}}

\newcommand{\PP}[4]{P_{#1#2}^{\phantom{#1#2}#3#4}}

\begin{document}

\begin{titlepage}
\vfill

\begin{flushright}
QMUL-PH-14-04
\end{flushright}

\vfill

\begin{center}
   \baselineskip=16pt
   	{\Large \bf Strings and Branes are Waves}
   	\vskip 2cm
   	{\sc		Joel Berkeley\footnote{\tt j.a.fitzhardinge-berkeley@qmul.ac.uk}, 
	 		David S. Berman\footnote{\tt d.s.berman@qmul.ac.uk} and
	 		Felix J. Rudolph\footnote{\tt f.j.rudolph@qmul.ac.uk}}
	\vskip .6cm
    {\small \it Queen Mary University of London, Centre for Research in String Theory, \\
             School of Physics, Mile End Road, London, E1 4NS, England} \\ 
	\vskip 2cm
\end{center}

\begin{abstract}
We examine the equations of motion of double field theory and the duality manifest form of M-theory. We show the solutions of the equations of motion corresponding to null waves correspond to strings or membranes from the usual spacetime perspective. A Goldstone mode analysis of the null wave solution in double field theory produces the equations of motion of the duality manifest string.
\end{abstract}

\vfill

\setcounter{footnote}{0}
\end{titlepage}

\tableofcontents

\section{Introduction}
One of the attractive elements of Kaluza-Klein theory is that it provides a single geometric construction for the Maxwell field and its action. The price is that we need to envoke an extra dimension and, if we wish to not have a whole new tower of massive states, we must also insist that fields are independent of this new dimension. We can also ask what is the interpretation, from the reduced spacetime point of view, of fields with a dependence on the KK coordinate. These are states charged with respect to the KK gauge field, with the charge being related to the momentum in the KK direction. These states will be massive from the reduced perspective since the momentum in the compact space will also appear as a mass from this perspective. Having a massless state in the five-dimensional theory with momentum along the fifth direction will then lead to a BPS state in the reduced four-dimensional theory as its charge will equal its mass (in four-dimensional) natural units.  

The identification of states whose mass and charge have their origin from KK momentum was crucial in the identification of the low energy effective action of M-theory. The D0-brane was simply a momentum mode along the eleventh direction \cite{Townsend:1995kk,Witten:1995ex}. Explicitly, from the eleven-dimensional supergravity perspective it was a null wave solution. From the reduced ten-dimensional perspective this could be identified with the D0-brane solution in IIA supergravity, its charge and mass originating from the eleven-dimensional momentum of the null wave \cite{Townsend:1995kk,Townsend:1997wg}.

 Double Field Theory or hence forth DFT \cite{Hull:2009mi} and its subsequent developments in \cite{Hull:2009zb, Hohm:2010jy, Hohm:2010pp, Jeon:2010rw, Jeon:2011cn, Jeon:2011vx, Jeon:2011sq, Jeon:2012hp, Aldazabal:2011nj, Coimbra:2011nw} (see \cite{Aldazabal:2013sca, Berman:2013eva, Hohm:2013bwa} for recent reviews), may be viewed as an attempt to geometrically unify the metric and NS-NS two-form potential $B_{[2]}$ in a Kaluza-Klein type way. Amongst other reasons, the local symmetry of the NS-NS two-form means that one cannot lift this to just ordinary Riemannian geometry in higher dimensions. Instead one needs to have a so called {\it{generalized geometry}}. Double field theory extends the dimensions of spacetime so that the off-diagonal components of the generalized metric---that is the metric of the full extended space---become the NS-NS two-form potential. Then one solves the so-called {\it{strong constraint}} or {\it{section condition}}, that means effectively one then carries out a Kaluza-Klein reduction from the full extended space down to usual spacetime. The action of DFT then reduces to the ordinary supergravity action. The generalized diffeomorphisms become both the ordinary diffeomorphisms and two-form gauge transformations. (The global aspects of which have recently been explored in \cite{Hohm:2012mf, Berman:2014jba, Cederwall:2014kxa}.) As such we can view DFT as a novel type of Kaluza-Klein theory which lifts the NS-NS sector of supergravity (i.e. metric and two-form) to a single geometric theory in higher dimensions. 

The extended geometry associated with the duality manifest version of M-theory \cite{Hillmann:2009ci, Hull:2007zu, Pacheco:2008ps, Coimbra:2011ky, Coimbra:2012af, Berman:2010is, Berman:2011pe, Berman:2011jh} is a further extension of this idea where the three-form potential $C_{[3]}$ and the metric are combined and lifted into a {\it{generalized metric}} for a single geometric theory with an extended number of dimensions. Again there is a section condition \cite{Berman:2011cg, Coimbra:2011ky, Coimbra:2012af, Berman:2012vc} whose solution implies a Kaluza-Klein reduction back to ordinary spacetime.

It is natural to ask the question what is the interpretation of momenta along the extra directions. A few moments thought about the comparision between DFT and Kaluza-Klein theory indicates that it should correspond to a fundamental string charge. Thus this indicates an intruiging interpretation for the fundamental string from the DFT point of view. The string will just be a null wave in doubled space with the momentum along the extended directions. The $O(d,d)$ symmetry of T-duality which from the usual spacetime point of view exchanges winding and momentum will now just correspond to a rotation in the doubled space. A null wave pointing along the usual spacetime will be a momentum mode but pointing along an extended direction it will be interpreted as a fundamental string. The charge and tension of the string will just be given by the momentum. Thus from the DFT point of view there are no strings, only null waves.

We will make this connection as explicit as possible. We begin by constructing a null pp-wave solution of the equations of motion of DFT and interpret it as a massless state in doubled space carrying momentum. We then show that this is the fundamental string solution \cite{Dabholkar:1990yf} when written in terms of the usual spacetime metric and two-form potential. We wish to study the dynamics of such a solution. To do so we determine the equations of motion of the Goldstone modes of this null wave solution in DFT. (Technically we follow \cite{Adawi:1998ta} very closely.) The resulting equations of motion for the Golsstone modes are the same as that of the string theory written down by Tseytlin \cite{Tseytlin90, Tseytlin91} to describe a string world-sheet in doubled space. 

We then move to exhibit the same property for the duality manifest form of M-theory (with U-duality group $SL(5)$). The wave is shown to be equivalent to the membrane. Thus again there are no fundamental extended objects, only null waves. Along the way we will need to write down the equations of motion of the duality manifest theory---something that has so far not been done. Even though the action for the manifest $SL(5)$ theory has been known for a few years by now \cite{Berman:2010is}, the equations of motion are more complicated than the Euler-Lagrange equations from that action since the generalized metric is constrained to be an element of the $SL(5)/SO(5)$ coset. Implementing this constraint in the variational problem of the action then leads to a projected set of equations of motion just as in \cite{Hohm:2010pp} for DFT. We then conjecture the general form for the projector in terms of the Y-tensor introduced in \cite{Berman:2012vc}.

\subsection{Bibliography}
It is out of the scope of this paper to give a proper historical account of DFT and its development. There are three relatively recent reviews of the subject \cite{Aldazabal:2013sca, Berman:2013eva, Hohm:2013bwa}. We would like to emphasize the early work of Siegel \cite{Siegel93a, Siegel93b} and Duff \cite{Duff90a} and then the two key groups that have developed DFT, one of Hohm, Hull and Zwiebach \cite{Hull:2009mi, Hull:2009zb, Hohm:2010jy, Hohm:2010pp} and the other of Jeon, Lee and Park \cite{Jeon:2010rw, Jeon:2011cn, Jeon:2011vx, Jeon:2011sq, Jeon:2012hp}. In the duality manifest M-theory formalism there was initial work by Duff \cite{Duff90b} and then Hull \cite{Hull:2007zu} and Waldram et. al. \cite{Pacheco:2008ps, Coimbra:2011ky, Coimbra:2012af} and later, Berman, Perry and collaborators \cite{Berman:2010is, Berman:2011pe, Berman:2011cg, Berman:2011jh}. Recently some key further developments in this direction are by Grana et. al. \cite{Aldazabal:2013via} and Hohm and Sambtleben \cite{Hohm:2013pua, Hohm:2013vpa, Hohm:2013uia}. From one perspective many of these developments were anticipated by the so called $E_{11}$ programme of West and collaborators \cite{West:2001as, Englert:2003zs, West:2003fc, Kleinschmidt:2003jf, West:2004kb, West:2012qm}. As such many of the ideas present in DFT and its variants were signalled by the early work of West. In particular the authors of this paper have been influenced by the fact that the nonlinear realization construction central to the $E_{11}$ programme has its origins in the theory of pions as Goldstone modes of the spontanteously broken chiral Lagrangian. This led to the idea that the duality invariant theory may contain massless Goldstone modes from spontaneously breaking the duality symmetry. Whether the null states identifed here are such Goldstone modes is an open question. 

For quantum aspects of the duality manifest string see \cite{Berman:2007vi,Berman:2007xn,Berman:2007yf,Hohm:2013jaa,Betz:2014aia}. In addition, there have been a whole host of fascinating recent results, some small sample of which are \cite{Berman:2013uda, Blair:2013noa, Blair:2013gqa, Lee:2014mla, Lee:2013hma,Strickland-Constable:2013xta, Park:2014una, Cederwall:2013naa}. When studying supergravity solutions such as the pp-wave, the string, the membrane and the D0-brane, as well as reviewing concepts like T-duality, Kaluza-Klein reductions and smearing we found the book by Ortin \cite{Ortin04} an invaluable reference.

\subsection{Notation}
In this paper we are dealing with several different spaces of various dimensions at the same time. Here is a brief summary of the indices and their ranges used for these spaces. We start with the spacetime of dimension $d$ with metric $g_{\mu\nu}$ and coordinates $x^\mu$ where $\mu=1,\dots,d$. In DFT this is the normal $d$-dimensional space and for the $SL(5)$ duality invariant theory where the dimensions are split into 4+7, these are the four dimensions the U-duality group acts on, thus $d=4$.

The {\it{duals}} of the spacetime coordinates are denoted by $\tx_\mu$ or $\tx^\bmu$ for DFT and $y_{\mu\nu}$ for the $SL(5)$ theory. Together with the normal coordinates $x^\mu$ they form a doubled or extended space of dimension $D$ with coordinates $X^M$ and generalized metric $\HH_{MN}$ for DFT and $\MM_{MN}$ for the $SL(5)$ theory, where $M=1,\dots,D$. In DFT we have $D=2d$ and the doubled space is equipped with an $O(d,d)$ structure. In $SL(5)$ there are six wrapping coordinates $y_{\mu \nu}$, where $\mu, \nu$ are antisymmetrized and thus $D=10$.

In what follows, we will see that the equations of motion will be projected using a projector denoted by  ${P_{MN}}^{KL}$. This acts on a $(D\times D)$-dimensional symmetric vector space whose building blocks are ``vectors'' of the form $V_{MN}$ with $M,N$ symmetrized. The dimension of this vector space is therefore $\frac{1}{2}D(D+1)$.

All the dimension and indices are summarized in the following table.
\begin{equation}
\begin{array}{|l|c|ccc|c|}
\hline
\mathrm{space}		& \mathrm{dimension}	&	O(d,d)	& SL(5)	&	SO(5,5)	& \mathrm{indices} \\\hline
\mathrm{spacetime}	& 	d				&	d		&	4	&	5		& \mu,\nu,\dots	\\
\mathrm{extended\ space}	& 	D				&	2d		&	10	&	16		& M,N,\dots \\
\mathrm{projector\ space}	& \frac{1}{2}D(D+1) 	& 2d^2+d		& 	55	& 	136		& (MN), (PQ), \dots \\ \hline
\end{array}
\end{equation}

\subsection{Double Field Theory}
\label{sec:DFTintro}
In double field theory the spacetime metric $g_{\mu\nu}$, the B-field $B_{\mu\nu}$ and the dilaton $\phi$ are encoded in the generalized metric $\HH_{MN}$ and the rescaled dilation $d$ as follows,
\begin{align}
\HH_{MN} &= 
\begin{pmatrix}
g_{\mu\nu} - B_{\mu\rho}g^{\rho\sigma}B_{\sigma\nu} & B_{\mu\rho}g^{\rho\nu} \\
-g^{\mu\sigma}B_{\sigma\nu} & g^{\mu\nu} 
\end{pmatrix}
\qandq
d= \phi - \frac{1}{4}\ln g
\label{eq:DFTmetric}
\end{align}
where $g=\det g_{\mu\nu}$ is the determinant of the spacetime metric. This generalized metric is then a metric on a $2d$ dimensional space. We introduce the usual coordinates $x^\mu$ and their duals $\tx_\mu$ which are combined into $X^M=(x^\mu, \tx_\mu)$ for the whole doubled space. This doubled space is also equipped with a globally defined $O(d,d)$ structure $\eta_{MN}$
\begin{equation}
\eta_{MN} =
\begin{pmatrix}
0 & {\delta_\mu}^\nu \\ {\delta^\mu}_\nu & 0
\end{pmatrix}
\label{eq:eta}
\end{equation}
and all tensors are really $O(d,d)$ tensors in the doubled space (for a discussion of this see \cite{Berman:2014jba}). The action may then be written in terms of a sort of generalized Ricci scalar
\begin{equation}
S = \int \dd^{D}X e^{-2d} R
\label{eq:DFTaction}
\end{equation}
with the scalar $R$ given by
\begin{equation}
\begin{aligned}
R 	&= \frac{1}{8}\HH^{MN}\partial_M\HH^{KL}\partial_N\HH_{KL} 
		- \frac{1}{2}\HH^{MN}\partial_M\HH^{KL}\partial_K\HH_{NL} \\
	&\quad+ 4\HH^{MN}\partial_M\partial_N d - \partial_M\partial_N\HH^{MN}
		-4\HH^{MN}\partial_M d \partial_N d + 4\partial_M\HH^{MN}\partial_N d \\
	&\quad+ \frac{1}{2}\eta^{MN}\eta^{KL}\partial_M{\EE^A}_K\partial_N{\EE^B}_L\HH_{AB} \, . 
\end{aligned}
\end{equation}
Here $M,N$ are \emph{curved} doubled spacetime indices and $A,B$ are \emph{flat} doubled tangent space indices. They are related via the generalized vielbeins ${\EE^A}_M$ such that
\begin{equation}
\HH_{MN} = {\EE^A}_M{\EE^B}_N \eta_{AB} \, .
\end{equation}
In addition to this there is the so called section condition or strong constraint. This diminishes the dependence of the fields on the number of coordinates. This constraint may be written as
\begin{equation}
\eta^{MN}\partial_M  \bullet  \partial_N \bullet =0
\label{eq:sectioncond}
\end{equation}
for any field in the theory. Its simple consequence is that one may choose to have dependence on the usual coordinates alone. Different choices of how one solves the section condition gives rise to different duality related theories. So at the cost of breaking the $O(d,d)$ symmetry we may choose
\begin{equation}
\partial_\bmu    \bullet =0  \, .  
\label{eq:sectioncondKK}
\end{equation}
This is like a simple Kaluza-Klein reduction and we will find it useful in what follows to take this perspective. Imposing the condition \eqref{eq:sectioncondKK} on the action \eqref{eq:DFTaction} produces the NS-NS sector of supergravity. (There is also a boundary term contribution that will not play a role in what follows \cite{Berman:2011kg}.) Thus at a rather simplistic level, the DFT action is like a Kaluza-Klein lift of the NS-NS sector of supergravity. Note, the last line in $R$ containing the vielbeins was originally not present in the literature. This is because indeed it vanishes when one imposes the section condition. It is however crucial when one considers the Scherk-Schwarz reductions of the theory \cite{Berman:2012uy, Berman:2013cli, Grana:2012rr, Geissbuhler:2011mx, Aldazabal:2013mya}. (We will not consider such Scherk-Schwarz reductions in this paper.)

The equation of motion for the dilaton is easily obtained by varying the action
\begin{equation}
\delta S = \int \dd^{D}X e^{-2d} (-2R)\delta d
\end{equation}
which has to vanish for any variation $\delta d$ and thus gives
\begin{equation}
R=0
\label{eq:DFTeomDilaton}.
\end{equation}
(Note that $\delta R/\delta d=0$ up to total derivatives.) To find the equation of motion for the generalized metric we have to be a bit more careful. Varying the action with the generalized metric gives
\begin{equation}
\delta S = \int \dd^{D}X e^{-2d} K_{MN}\delta \HH^{MN}
\label{eq:DFTvaraction}
\end{equation}
where $K_{MN}$ is  given by
\begin{equation}
\begin{aligned}
K_{MN} &= \frac{1}{8}\partial_M\HH^{KL}\partial_N\HH_{KL} + 2\partial_M\partial_N d \\
	&\quad +(\partial_L-2\partial_L d) 		
		\left[\HH^{KL}\left(\partial_{(M}\HH_{N)K}
		- \frac{1}{4}\partial_K\HH_{MN}\right)\right] \\
	&\quad  + \frac{1}{4}\left(\HH^{KL}\HH^{PQ}-2\HH^{KQ}\HH^{LP}\right)
		\partial_K\HH_{MP}\partial_L\HH_{NQ} \\
	&\quad - \eta^{KL}\eta^{PQ}\left(\partial_K d\partial_L{\EE^A}_P 
		- \frac{1}{2}\partial_K\partial_L{\EE^A}_P\right)\HH_{(N|R}{\EE^R}_A\HH_{|M)Q} \, . 
\end{aligned}
\end{equation}
The last term uses the variation of the vielbein with respect to the metric
\begin{equation}
\delta{\EE^A}_M = \frac{1}{2}\HH^{AB}{\EE^N}_B\delta\HH_{MN} \, .
\end{equation} 

The expression in \eqref{eq:DFTvaraction} does not have to vanish for any $\delta\HH^{MN}$ since the generalized metric is constrained to parametrize the coset space $O(d,d)/O(d)\times O(d)$. This means the generalized metric can be parametrized by $g_{\mu\nu}$ and $B_{\mu\nu}$ as written in \eqref{eq:DFTmetric}. Thus deriving the equations of motion is a little more complicated. This was first done in \cite{Hohm:2010pp}. We will rederive the equations of motion here using a slightly different method because  this method will be more readily applicable to the cases of extended geometry with the exceptional groups that we discuss later. The basic idea is that rather than varying with respect to the generalized metric one varies with respect to the spacetime metric and the B-field and then make the result $O(d,d)$ covariant.

By applying the chain rule, the action can be varied with respect to $\delta g_{\mu\nu}$ and $\delta B_{\mu\nu}$ separately. Making use of
\begin{align}
\frac{\delta g_{\mu\nu}}{\delta g_{\rho\sigma}} &= {\delta_\mu}^{(\rho}{\delta_\nu}^{\sigma)}, &
\frac{\delta g^{\mu\nu}}{\delta g_{\rho\sigma}} &= -g^{\mu(\rho}g^{\sigma)\nu}, &
\frac{\delta B_{\mu\nu}}{\delta B_{\rho\sigma}} &= {\delta_\mu}^{[\rho}{\delta_\nu}^{\sigma]}
\end{align}
leads to
\begin{align}
\delta S &= \int \dd^{D}X e^{-2d} K_{MN}\left[
		\frac{\delta \HH^{MN}}{\delta g_{\rho\sigma}}\delta g_{\rho\sigma}
		+ \frac{\delta \HH^{MN}}{\delta B_{\rho\sigma}}\delta B_{\rho\sigma}\right] \\
		&= \int \dd^{D}X e^{-2d} \left\{
			\left[- K_{\mu\nu}g^{\mu(\rho}g^{\sigma)\nu}
					+ 2{K_\mu}^\nu g^{\mu(\rho}g^{\sigma)\tau}B_{\tau\nu} 
					\vphantom{\left({\delta_\mu}^{(\rho}\right)}\right.\right. \notag\\
		& \left.\left. \hspace{4cm} 	+ K^{\mu\nu}\left({\delta_\mu}^{(\rho}{\delta_\nu}^{\sigma)} 
					+ B_{\mu\tau}g^{\tau(\rho}g^{\sigma)\lambda}B_{\lambda\nu}
					\right)\right]\delta g_{\rho\sigma} \right. \\
		& \left. \hspace{3.5cm} + 
			\left[- 2{K_\mu}^\nu g^{\mu\tau}{\delta_\tau}^{[\rho}{\delta_\nu}^{\sigma]}
					- 2K^{\mu\nu}B_{\mu\tau}g^{\tau\lambda}
						{\delta_\lambda}^{[\rho}{\delta_\nu}^{\sigma]} \right]\delta B_{\rho\sigma} 	
			\right\}\, .\notag
\end{align}
Now the $g$'s and $B$'s are re-expressed in terms of $\HH$, the symmetrizing brackets are dropped and the antisymmetrizing ones are expanded
\begin{align}
\delta S &= \int \dd^{D}X e^{-2d} \left\{\vphantom{\frac{1}{2}}
	\left[- K_{\mu\nu}\HH^{\mu\rho}\HH^{\sigma\nu}
			+ 2{K_\mu}^\nu \HH^{\mu\rho}{\HH^{\sigma}}_\nu 
			+ K^{\mu\nu}\left({\delta_\mu}^{\rho}{\delta_\nu}^{\sigma} 
					- {\HH_\mu}^{\rho}{\HH^{\sigma}}_\nu\right)\right]\delta g_{\rho\sigma}\right. \notag\\
		& \left. \hspace{3.5cm} -2 
			\left[{K_\mu}^\nu \HH^{\mu\tau} + K^{\mu\nu}{\HH_\mu}^\tau\right]
			\frac{1}{2}\left({\delta_\tau}^{\rho}{\delta_\nu}^{\sigma} 
				- {\delta_\tau}^{\sigma}{\delta_\nu}^{\rho}\right) \delta B_{\rho\sigma} 	
			\right\}.
\end{align}
The crucial step is to then re-covariantize the indices by using $\eta_{MN}$ given in \eqref{eq:eta}
\begin{equation}
\begin{aligned}
\delta S &= \int \dd^{D}X e^{-2d} \left\{
	K_{KL} \left(\eta^{K\rho}\eta^{\sigma L} - \HH^{K\rho}\HH^{\sigma L}\right) 
		\delta g_{\rho\sigma}\right. \\
	& \left. \hspace{3cm} - K_{KL}\left(\HH^{KP}\eta_{PM}\eta^{LN} - \HH^{KP}{\delta_P}^N{\delta_M}^L\right)
	\eta^{M\rho}{\delta^\sigma}_N\delta B_{\rho\sigma} \right\} 
\end{aligned}
\end{equation}
which reproduces the previous line once the doubled indices are expanded and summed over. In a final step the terms inside the brackets are brought into a form corresponding to a projected set of equations as follows
\begin{align}
\delta S &= \int \dd^{D}X e^{-2d} \left\{
	K_{KL} \left({\delta_M}^K{\delta_N}^L - \HH^{KP}\eta_{PM}\eta_{NQ}\HH^{QL}\right) 	
		\eta^{M\rho}\eta^{\sigma N}\delta g_{\rho\sigma}\right. \notag\\
	& \left. \hspace{3cm} - K_{KL}\left(\HH^{KP}\eta_{PM}\eta^{LQ}\HH_{QR} - \HH^{KP}{\delta_P}^Q{\delta_M}^L\HH_{QR}\right)
	\HH^{RN}\eta^{M\rho}{\delta^\sigma}_N\delta B_{\rho\sigma} \right\} \notag\\
	&= \int \dd^{D}X e^{-2d} 2{P_{MN}}^{KL}K_{KL}
		\left(\eta^{M\rho}\eta^{\sigma N}\delta g_{\rho\sigma}
			+ \eta^{M\rho}\HH^{\sigma N}\delta B_{\rho\sigma}\right)
\end{align}
where we have introduced the projector
\begin{equation}
{P_{MN}}^{KL} = \frac{1}{2}({\delta_M}^{(K}{\delta_N}^{L)} 
	- \HH_{MP}\eta^{P(K}\eta_{NQ}\HH^{L)Q})
\end{equation}
which is symmetric in both $MN$ and $KL$. 

The variation of the action has to vanish for \emph{any} $\delta g_{\mu\nu}$ and $\delta B_{\mu\nu}$ independently, therefore the equations of motion are given by  
\begin{equation}
{P_{MN}}^{KL}K_{KL} = 0
\label{eq:DFTeom}
\end{equation}
and not $K_{MN}=0$, the naive equations expected from setting \eqref{eq:DFTvaraction} to zero. 

This equation of motion was derived in a slightly different way in \cite{Hohm:2010pp} by using the constraint equation $\HH^t\eta\HH=\eta$ which ensures $\HH$ is an element of $O(d,d)$. The result is 
\begin{equation}
\frac{1}{2}(K_{MN} - \eta_{MK}\HH^{KP}K_{PQ}\HH^{QL}\eta_{LN}) = {P_{MN}}^{KL}K_{KL} = 0
\end{equation}
in agreement with ours. We wish to emphasize the point of rederiving these equations is just so that we can use this method in the exceptional case later. 

Also note that the expression for $K_{MN}$ found in the literature, especially in \cite{Hohm:2010pp}, differs from the one given here. This difference arises as one can use either the invaraint $O(d,d)$ metric $\eta$ or the generalized metric $\HH$ to raise and lower indices in the derivation of $K_{MN}$. Both methods are valid and the discrepancy disappears once the projector acts. In a way, the projector enforces the constraint that $\HH$ parametrizes a coset space. When using $\eta$, this constraint is taken into account automatically, but when using $\HH$ the constraint needs to be imposed by the projector. Since $K_{MN}$ appears in the equations of motion only with the projector acting on it, it does not matter which version is used.

The importance for the presence of the projector can be seen by counting degrees of freedom. The symmetric spacetime metric has $\frac{1}{2}d(d+1)$ degrees of freedom and the antisymmetric B-field contributes $\frac{1}{2}d(d-1)$ for a total of $d^2$ independent components. The dimension of the doubled space is $D=2d$, therefore $K_{MN}$ has $2d^2+d$ components. Of these, $d^2+d$ are in the kernel of the projector and are therefore eliminated, leaving $d^2$ degrees of freedom as desired. This can be shown by computing the characteristic polynomial and all the eigenvalues of the projector $P$.

\section{The String as a Wave}
Now we are equipped with the equations of motion of DFT and so we move on to describe a solution of these equations and subsequently examine its Goldstone modes.

\subsection{Wave Solution or Fundamental String in DFT}
\label{sec:F1string}
We seek a solution for the generalized metric corresponding to a null wave whose momentum is pointing the $\tz$ direction. The ansatz will be that of a pp-wave in usual general relativity \cite{Aichelburg:1970dh}. This has no compunction to be a solution of DFT. As we have seen the equations of motion of the generalized metric in DFT are certainly not the same as the equations of motion of the metric in relativity. Let us immediately remove any source of confusion the reader may have, the pp-wave as a solution for $g_{\mu\nu}$ may of course, by construction, be embedded as a solution in DFT by simply inserting the pp-wave solution for $g_{\mu \nu}$ into $\HH_{MN}$. Here we will consider a pp-wave (that is the usual pp-wave ansatz \cite{Aichelburg:1970dh}) not for $g_{\mu\nu}$ but for the doubled metric $\HH_{MN}$  itself and then determine its interpretation in terms of the usual metric $g_{\mu\nu}$ and two-form $B_{\mu\nu}$.

The following is a solution to DFT in $2d$ dimensions given by the generalized metric $\HH_{MN}$ with line element
\begin{equation}
\begin{aligned}
\dd s^2 &= \HH_{MN}\dd X^M \dd X^N \\
	&= (H-2)\left[\dd t^2 - \dd z^2\right] + \delta_{mn}\dd y^m\dd y^n \\
	&\quad + 2(H-1)\left[\dd t\dd\tz + \dd\ttt\dd z\right] \\
	&\quad - H\left[\dd\ttt^2 - \dd\tz^2\right] + \delta^{mn}\dd\ty_m\dd\ty_n
\label{eq:DFTppwave}	
\end{aligned}
\end{equation}
where the generalized coordinates are split as
\begin{equation}
X^M = (x^\mu,\tx_\mu)=(t,z,y^m;\ttt,\tz,\ty_m)
\end{equation}
and a tilde denotes a dual coordinate as explained above. This generalized metric and rescaled dilaton $d=const.$ solve the equations of motion of the DFT derived in Section \ref{sec:DFTintro}. The appendix \ref{sec:DFTcheck} provides the details demonstrating it is indeed a solution.

Since it is exactly the same form as the usual pp-wave solution, the natural interpretation is of a pp-wave in the doubled geometry. One therefore imagines it propagates and therefore carries momentum in the $\tz$ direction. It is worth a pausing here. To determine whether it truely carries momentum would require the construction of conserved charges in DFT. This has not yet been done. It would be useful to consider objects like generalized Komar integrals  and the other ways one defines charges in general relativity but now for DFT. Nevertheless, we shall proceed with the interpretation of this solution as a pp-wave and thus carries momentum in the dual $\tz$ direction.

$H$ is taken to be a harmonic function of the usual transverse coordinates\footnote{The range of the transverse index is $m=1,\dots,d-2$.} $y^m$ (but not of their duals $\ty_m$) and as such is annihilated by the Laplacian operator in these directions, i.e. $\delta^{mn}\partial_m\partial_n H=0$. In DFT language, it is required (at least naively) that $H$ satifies the section condition and so to solve the section condition it is not a function of any of the dual coordinates. The fact that the harmonic function $H$ is taken to only depend on $y^m$ and not the dual transverse directions implies that the wave solution is {\it{smeared}} in these $\ty_m$ directions. One can think of it as a plane wave front extending along the dual directions described by coordinates $\ty_m$ but with momentum in the $\tz$ direction. An explicit form of $H$ is
\begin{equation}
H = 1 + \frac{h}{r^{d-4}} \qquad \mathrm{for} \qquad r^2 = y^my^n\delta_{mn}
\end{equation}
where $h$ is a constant and $r$ is the radial coordinate of the transverse space.

We will now use the form of the doubled metric $\HH_{MN}$ in terms of $g_{\mu\nu}$ and $B_{\mu\nu}$ to rewrite this solution in terms of $d$-dimensional quantities, effectively reducing the dual dimensions. This is like in Kaluza-Klein theory, writing a solution of the full theory in terms of the reduced metric and vector potential 
\begin{align}
\dd s^2 &= (g_{\mu\nu} - B_{\mu\rho}g^{\rho\sigma}B_{\sigma\nu})\dd x^\mu \dd x^\nu
	+ 2B_{\mu\rho}g^{\rho\nu}\dd x^\mu \dd\tx_\nu + g^{\mu\nu}\dd\tx_\mu\dd\tx_\nu \, .
\label{eq:KKforDFT}  
\end{align}

By comparing \eqref{eq:KKforDFT} with \eqref{eq:DFTppwave}, the fields of the reduced theory with coordinates $x^\mu=(t,z,y^m)$ can be computed. We find the metric and its inverse to be
\begin{equation}
g_{\mu\nu} = \diag (-H^{-1}, H^{-1}, \delta_{mn}) \qandq 
g^{\mu\nu} = \diag (-H, H, \delta^{mn})
\end{equation}
whereas the only non-zero component of the B-field is given by
\begin{equation}
B_{tz} = -B_{zt} = -(H^{-1}-1) \, .
\end{equation}
From the definition $e^{-2d} = \sqrt{g}e^{-2\phi}$ of the rescaled dilaton $d$ (which is a constant here) it follows that the dilaton $\phi$ is given by ($\phi_0$ is another constant)
\begin{equation}
e^{-2\phi} = H e^{-2\phi_0} \qquad \mathrm{or} \qquad e^{-2(\phi-\phi_0)} = H
\end{equation}
since $g=-H^{-2}$. The corresponding line element is 
\begin{equation}
\dd s^2 = -H^{-1}(\dd t^2-\dd z^2)+\delta_{mn}\dd y^m\dd y^n 
\label{eq:string}
\end{equation}
which together with the B-field and the dilaton $\phi$ gives the fundamental string solution extended along the $z$ direction \cite{Dabholkar:1990yf}. We have thus shown that the solution \eqref{eq:DFTppwave} which carries momentum in the $\tz$ direction in the doubled space corresponds to the string along the $z$ direction from a reduced point of view. 

This follows the logic of usual Kaluza Klein theory. In the doubled formalism the solution is a massless wave with $P_MP_N\HH^{MN}=0$ (where the $P^M$ are some generalized momenta), but from a the reduced normal spacetime point of view the string has a tension $T$ and charge $q$ which are obviously given by the momenta in the dual directions with a resulting BPS equation
\begin{equation}
T = |q| \, .
\end{equation}

Of course this is no surprise from the point of view of T-duality. Momentum  and string winding exchange under T-duality. It is precisely as expected that momentum in the dual direction corresponds to a string. What is more surprising is when one views this from the true DFT perspective. There are null wave solutions that can point in any direction. When we analyze these null waves from the reduced theory we see them as fundamental strings or as usual pp-waves. It is a simple $O(d,d)$ rotation of direction of propagation that takes one solution into the other. This is duality from the DFT perspective.

\subsection{Goldstone Modes of the Wave Solution}
\label{sec:DFTgoldstones}
In the previous section we presented a solution to the equations of motion of DFT which reduces to the fundamental string. It will be  interesting to analyse the Goldstone modes of this solution in double field theory. Especially since the advent of M-theory, it was understood that branes are dynamical objects and that when one finds a solution of the low energy effective action one can learn about the theory by examining the dynamics of the Goldstone modes. For D-branes in string theory this was done in \cite{Adawi:1998ta} and for the membrane and fivebrane in M-theory, where such an analysis was really the only way of describing brane dynamics, this was done in  \cite{Kaplan:1995cp, Adawi:1998ta}. We will follow the excellent exposition and the method described in \cite{Adawi:1998ta} as closely as possible.

In DFT, the diffeomorphisms and gauge transformations are combined into generalized diffeomorphisms generated by a generalized Lie derivative. We will consider small variations in the generalized metric, $h_{MN}$ and the dilaton, $\lambda$ generated by such transformations as follows
\begin{align}
h_{MN} &= \delta_\xi \HH_{MN} = \LL_\xi \HH_{MN} \, ,  &
\lambda &= \delta_\xi d = \LL_\xi d  \, .
\end{align}
For all the duality invariant geometries including DFT, the generalized Lie derivative of the metric \cite{Hull:2009zb} is given by the ordinary Lie derivative plus a  correction in terms of the so called Y-tensor
\begin{equation}
\begin{aligned}
\LL_\xi \HH_{MN} &= L_\xi \HH_{MN} 
		- {Y^{LP}}_{MQ}\partial_P\xi^Q\HH_{LN}
		- {Y^{LP}}_{NQ}\partial_P\xi^Q\HH_{ML} \\
	&=	\xi^L\partial_L \HH_{MN} 
		+ 2\HH_{L(M}\partial_{N)}\xi^L 
		- 2{Y^{LP}}_{Q(M}\HH_{N)L}\partial_P\xi^Q \, .
\end{aligned}
\label{eq:genLieMetric}
\end{equation}
The Y-tensor \cite{Berman:2012vc} encodes a great deal about the geometry. For DFT, the Y-tensor is simply given in terms of the $O(d,d)$ metric
\begin{equation}
{Y^{MN}}_{KL}=\eta^{MN} \eta_{KL}   \, .
\end{equation}
If the metric $\HH_{MN}$ and the transformation parameter $\xi^{M}=(\xi^\mu,\tilde{\xi}_\mu)$ both satisfy the section condition, then the vector part $\xi^\mu$ generates a coordinate transformation while the one-form part $\tilde{\xi}_\mu$ gives a gauge transformation of the B-field. 

The generalized Lie derivative of the dilaton contains just the transport term plus a term for $d$ being a tensor density
\begin{align}
\LL_\xi d &=	\xi^M\partial_M d - \frac{1}{2}\partial_{M}\xi^M \, .
\label{eq:genLieDilaton}
\end{align}

The wave solutions are extended objects and therefore sweep out a worldvolume in space. This is spanned by the coordinates $\{t,z\}$. All remaining coordinates are treated as transverse in the extended space. The solution clearly breaks translation symmetry and so one naturally expects scalar zero-modes. One immediate puzzle would be to ask about the number of degrees of freedom of the Goldstone modes. Given that the space is now doubled one would naively image that any solution which may be interpreted as a string would have $2d-2$ degrees of freedom rather than the expected $d-2$. We will answer this question and show how the Goldstone modes have the correct number of degrees of freedom despite the solution living in a $2d$ dimensional space. The projected form of the equations of motion are crucial in making this work out.

To carry out the analysis it will be useful to split up the space into parts longitudinal and transverse to the string. One collects the worldvolume coordinates $t$ and $z$ into $x^a$ and similarly for their duals\footnote{In what follows we will use the alternative notation $\tx^\bmu$ for the dual coordinates to avoid confusion between inverse and dual parts of the metric.} $\tx^\ba = (\ttt,\tz)$ such that the generalized coordinates are $X^M=(x^a,y^m,\tx^\ba,\ty^\bm)$. This allows the non-zero components of the metric and its inverse to be written as
\begin{equation}
\begin{aligned}
\HH_{ab} &= (2-H)\II_{ab} 					&	\HH^{ab} &= H\II^{ab} \\
\HH_{\ba\bb} &= H\II_{\ba\bb} 				&	\HH^{\ba\bb} &= (2-H)\II^{\ba\bb}\\
\HH_{a\bb} &= \HH_{\bb a} = (H-1)\JJ_{a\bb}	&	\HH^{a\bb} &= \HH^{\bb a} = (H-1)\JJ^{a\bb} \\
\HH_{mn} &= \delta_{mn}, \quad \HH_{\bm\bn} = \delta_{\bm\bn}		
					&	\HH^{mn} &= \delta^{mn}, \quad\HH^{\bm\bn} = \delta^{\bm\bn}
\end{aligned}
\end{equation}
where the constant symmetric $2\times 2$ matrices $\II$ and $\JJ$ are defined as
\begin{equation}
\II = \begin{pmatrix} -1 & 0 \\ 0 & 1 \end{pmatrix}
\qandq
\JJ = \begin{pmatrix} 0 & 1 \\ 1 & 0 \end{pmatrix} \, .
\label{eq:IJmatrix}
\end{equation}
For later use also define their (antisymmetric) product
\begin{equation}
\KKK = \II \cdot \JJ = - \JJ\cdot\II = \begin{pmatrix} 0 & -1 \\ 1 & 0 \end{pmatrix} \, .
\label{eq:Kmatrix}
\end{equation}

Following \cite{Adawi:1998ta}, we now pick a transformation parameter $\xi^M$ with non-zero components only in the transverse directions, but with no transformation along the worldvolume directions (and the directions dual to the worldvolume). This transformation may then be described by the DFT vector field
\begin{equation}
\xi^M = (0,H^\alpha\hphi^m, 0, H^\beta\htphi^\bm)
\end{equation}
where $\hphi^m$ and $\htphi^\bm$ are the constant vectors that later will become the Goldstone modes once we allow them to have dependence on the worldvolume coordinates, $H$ is the harmonic function given above and $\alpha,\beta$ are constants that are to be determined by demanding that the Goldstone modes become normalisable. Using
\begin{equation}
h_{MN} = \xi^L\partial_L \HH_{MN} + 2\HH_{L(M}\partial_{N)}\xi^L 
		- 2\eta^{LP}\eta_{Q(M}\HH_{N)L}\partial_P\xi^Q
\end{equation}
we can compute the components of $h_{MN}$ in terms of $\hphi^m,\htphi^\bm$. Recall that both the metric and the transformation parameter only depend on $y$ through the harmonic function $H$. Therefore $\partial_m$ is the only derivative that gives a non-zero contribution. We find
\begin{equation}
\begin{aligned}
h_{ab} &= -\hphi^m (H^\alpha\partial_m H)  \II_{ab} 
	& h_{mn} &= 2\hphi^q\delta_{q(m}{\delta_{n)}}^p\partial_pH^\alpha \\
h_{\ba\bb} &= \hphi^m (H^\alpha\partial_m H)  \II_{\ba\bb} 
	& h_{\bm\bn} &= -2\hphi^q\delta_{q(\bm}{\delta_{\bn)}}^p\partial_p H^\alpha \\
h_{a\bb} &= h_{\bb a} = \hphi^m (H^\alpha\partial_m H)  \JJ_{a\bb}
	& h_{m\bn} &= h_{\bn m} = -2\htphi^\bq\delta_{\bq[m}{\delta_{\bn]}}^p\partial_p H^\beta
\end{aligned}
\label{eq:h}
\end{equation}
and all terms with indicies mixing $a,\ba$ with $m,\bm$ vanish. For the dilaton there is no contribution from the transport term as $d$ is a constant for our solution. This leaves the density term which gives
\begin{equation}
\lambda =  - \frac{1}{2}\hphi^m\partial_{m} H^\alpha \, .
\label{eq:lambda}
\end{equation}

Once we have these equations, the next step is to allow the moduli to have dependence on the worldvolume coordinates, 
\begin{equation}
\hphi^m\rightarrow\phi^m(x) \, , \qquad \htphi^\bm\rightarrow \tphi^\bm(x)  \label{eq:zeromodes}
\end{equation}
 and the hats are removed. These are the zero-modes.

We now determine their equations of motion by inserting \eqref{eq:zeromodes} into \eqref{eq:h} and \eqref{eq:lambda} and then subsequently into the equations of motion for DFT, \eqref{eq:DFTeomDilaton} and \eqref{eq:DFTeom}. As usual we keep only terms with two derivatives and first order in $h_{MN}$ and $\lambda$ themselves. (It would certainly be interesting to move beyond this expansion and compare with a Nambu-Goto type action but we will not do so here.) This gives
\begin{align}
K_{MN} &= \HH^{LK}\partial_L\partial_{(M} h_{N)K} 	
	- \frac{1}{4}\HH^{LK}\partial_L\partial_K h_{MN}
	+ 2\partial_M\partial_N \lambda \\
R &= 4 \HH^{MN}\partial_M\partial_N \lambda
		- \partial_M\partial_N h^{MN}.
\end{align}
For convenience we will define $\Box = H\II^{ab}\partial_a\partial_b$ and $\Delta = \delta^{kl}\partial_k\partial_l$. Inserting $h_{MN}$ from \eqref{eq:h}, we find
\begin{equation}
\begin{aligned}
K_{ab} &= -(1+\alpha H^{-1})\partial_a\partial_b\phi^m (H^\alpha\partial_m H) 
	+ \frac{1}{4}\II_{ab}\Box\phi^m(H^\alpha\partial_m H) \\
K_{\ba\bb} &= -\frac{1}{4}\II_{\ba\bb}\Box\phi^m(H^\alpha\partial_m H) \\
K_{a\bb} &= K_{\bb a} = \frac{1}{2}{\KKK^c}_\bb\partial_c\partial_a\phi^m 	
	(H^\alpha\partial_m H) - \frac{1}{4}\JJ_{a\bb}\Box\phi^m(H^\alpha\partial_m H) \\	
K_{mn} &= - \frac{\alpha}{2}\Box\phi^p\delta_{p(m}{\delta_{n)}}^q(H^\alpha\partial_qH) \\
K_{\bm\bn} &= \frac{\alpha}{2}\Box\phi^p\delta_{p(\bm}{\delta_{\bn)}}^q 
	(H^\alpha\partial_qH) \\
K_{m\bn} &= \delta K_{\bn m} = \frac{\beta}{2}\Box\tphi^\bp
	\delta_{\bp[m}{\delta_{\bn]}}^q(H^\beta\partial_q H) \\
K_{am} &= K_{ma} = \frac{1}{2}\partial_a\phi^n\left[
	\delta_{mn}\Delta H^\alpha - \partial_m\partial_n H^\alpha 
	- \partial_m(H^\alpha\partial_n H)\right] \\
K_{\ba m} &= K_{m\ba} = \frac{1}{2}{\KKK^b}_\ba\partial_b\phi^n
	\partial_m(H^\alpha\partial_n H) \\
K_{a\bm} &= K_{\bm a} = \frac{1}{2}\partial_a\tphi^\bn{\delta_\bn}^k{\delta_\bm}^l
	\left[\delta_{kl}\Delta H^\beta - \partial_k\partial_l H^\beta \right]\\
K_{\ba\bm} &= K_{\bm\ba} = 0
\end{aligned}
\end{equation}
where $\KKK$ was defined in \eqref{eq:Kmatrix}. Further, inserting $\lambda$ from \eqref{eq:lambda} gives the dilaton equation
\begin{equation}
R = -H^{-1}(2\alpha + 1) \Box\phi^m (H^\alpha\partial_m H) = 0.
\end{equation}
It is straight forward to see that the dilaton equation is solved by $\Box\phi = 0$. For the other equations we have to work a bit harder. The full equations of motion for the generalized metric are the projected equations \eqref{eq:DFTeom} which contain $d^2$ linearly independet equations
\begin{equation}
\begin{aligned}
K_{mn} &= {\delta_m}^\bk {\delta_n}^\bl K_{\bk\bl}  \\
K_{m\bn} &= {\delta_m}^\bk {\delta_\bn}^l K_{\bk l}
\end{aligned}
\label{eq:DFTblock1}
\end{equation}
\begin{equation}
\begin{aligned}
K_{mt} &= (H-1){\delta_m}^\bn K_{\bn z} - (2-H){\delta_m}^\bn K_{\bn\bt} \\
K_{mz} &= (H-1){\delta_m}^\bn K_{\bn t} + (2-H){\delta_m}^\bn K_{\bn\bz} \\
K_{m\bt} &= (H-1){\delta_m}^\bn K_{\bn\bz} - H{\delta_m}^\bn K_{\bn t} \\
K_{m\bz} &= (H-1){\delta_m}^\bn K_{\bn\bt} + H{\delta_m}^\bn K_{\bn z}
\end{aligned}
\label{eq:DFTblock2}
\end{equation}
\begin{equation}
\begin{aligned}
0 &= (H-1)(K_{\bt\bt} - K_{\bz\bz}) + H(K_{t\bz} + K_{z\bt}) \\
0 &= (H-1)(K_{tt} - K_{zz}) + (2-H)(K_{t\bz} + K_{z\bt}) \\
0 &= (H-1)(K_{t\bz} - K_{z\bt}) - HK_{zz} + (2-H)K_{\bz\bz} \\
0 &= (H-1)(K_{t\bt} - K_{z\bz}) + HK_{tz} + (2-H)K_{\bt\bz}.
\end{aligned}
\label{eq:DFTblock3}
\end{equation}
Inserting for $K_{MN}$ from above yields the equations of motion for the zero modes. The first two read
\begin{equation}
\begin{aligned}
-\alpha\Box\phi^p\delta_{p(m}{\delta_{n)}}^q(H^\alpha\partial_qH) &= 0 \\
\beta\Box\tphi^\bq\delta_{\bq[m}{\delta_{\bn]}}^p(H^\beta\partial_q H) &= 0 
\end{aligned}
\end{equation}
and can be solved by $\Box\phi=0$ and $\Box\tphi=0$ respectively. The next block of equations \eqref{eq:DFTblock2} can be re-covariantized by using
\begin{equation}
-\II_{ac}\epsilon^{cb}=
-\begin{pmatrix} -1 & 0 \\ 0 & 1 \end{pmatrix}\begin{pmatrix} 0 & 1 \\ -1 & 0 \end{pmatrix}
=\begin{pmatrix}0 & 1 \\ 1 & 0 \end{pmatrix}
\end{equation}
which leads to
\begin{equation}
\begin{aligned}
\partial_a\phi^n\left[\delta_{mn}\Delta H^\alpha - \partial_m\partial_n H^\alpha 
	\right. & \left.- \partial_m(H^\alpha\partial_n H)\right] \\
	&= -\II_{ac}\epsilon^{cb}\partial_b\tphi^\bn{\delta_\bn}^p
	(H-1)\left[\delta_{pm}\Delta H^\beta - \partial_p\partial_m H^\beta \right] \\
\partial_a\phi^n\partial_m(H^\alpha\partial_nH) 
	&= \II_{ac}\epsilon^{cb}\partial_b\tphi^\bn{\delta_\bn}^p
	H\left[\delta_{pm}\Delta H^\beta - \partial_p\partial_m H^\beta \right].
\end{aligned}
\end{equation}
Adding these two equations gives
\begin{equation}
\partial_a\phi^n W_{mn}^{(\alpha)} =  \II_{ac}\epsilon^{cb}\partial_b\tphi^\bn{\delta_\bn}^n W_{mn}^{(\beta)}
\end{equation}
where for $\gamma=\alpha,\beta$ we have $W_{mn}^{(\gamma)}=\delta_{mn}\Delta H^\gamma - \partial_m\partial_n H^\gamma$. If $\alpha=\beta$ we have the same object $W_{mn}$ on both sides which can be shown to be invertible. The equation can thus be reduced to a duality relation between $\phi$ and $\tphi$
\begin{equation}
\partial_a\phi^m = \II_{ab}\epsilon^{bc}\partial_c\tphi^\bn\delta_\bn^m 
\qquad \mathrm{or} \qquad
\dd\phi^m = \star\dd\tphi^\bn\delta_\bn^m.  
\label{eq:dualityzeromodes}
\end{equation}
This equation implies both $\Box\phi=0$ and $\Box\tphi=0$ as can be seen by acting with a contracted derivative on the equation. If $\phi^m$ and $\tphi^\bm$ are placed in a generalized vector $\Phi^M=(0,\phi^m,0,\tphi^\bm)$ this can be written as a self-duality relation
\begin{equation}
\HH_{MN}\dd\Phi^M = \eta_{MN}\star\dd\Phi^N 
\end{equation}
and precisely matches the result in \cite{Duff90a} for the duality symmetric string.

The final block of equations of motion \eqref{eq:DFTblock3} are either trivial or are also of the form $\Box\phi^m(H^\alpha\partial_mH)=0$ provided $\alpha=-1$. If one was not concerned by normalisation of the modes then this also provides a way of constraining the value of $\alpha$. The consistent choice of $\alpha=-1$ is fortunately the choice that also leads to normalisable modes. This may be seen by examing the case $\alpha=-1$ and integrating over the transverse space. This exactly mirrors the situation described in \cite{Adawi:1998ta}. The Goldstone modes are really the normalisable modes corresponding to broken gauge transformations. Where for gravity the gauge transformations are ordinary diffeomorphisms, in the case of DFT it is generated by the generalised Lie deriviative. (In case the reader is more familiar with the study of monopoles, the analogue of the modes described in this paper is with the dyonic $U(1)$ mode in the monopole moduli space.)

One can now turn equation \eqref{eq:dualityzeromodes} into a (anti-)chiral equation for a linear combination of $\phi$ and  $\tphi$ as follows. Introducing $\psi_\pm$ to be given by
\begin{equation}
\psi_\pm = \phi \pm \tphi
\end{equation}
and inserting them into \eqref{eq:dualityzeromodes} and its Hodge dual gives the familiar (anti-)self-dual left- and right-movers
\begin{equation}
\dd\psi_\pm = \pm\star\dd\psi_\pm 
\end{equation}
of the Tseytlin-string \cite{Tseytlin90, Tseytlin91}. Thus the dynamics of the Goldstone modes of the wave solution reproduce the duality symmetric string in doubled space. The number of physical degrees of freedom are not doubled but just become rearranged in terms of chiral and anti-chiral modes on the world-sheet.

\subsection{Comparison with the $\sigma$-model evaluated in the String or Wave Background}

The equations of motion that were derived in the previous section recover the equations of motion of the Tseytlin string. A natural question would be to ask what background is the string in? Is the target space of the doubled solution the combination of the fundamental string with the wave background? The answer to this question can be seen immediately from the Goldstone mode analysis which gives the equations of motion of the free string i.e. that of the $\sigma$-model in a flat background. 

To understand this it is worth understanding what the Goldstone mode analysis provides you with in other cases where this has been carried out in a more conventional setting. In the work of \cite{Adawi:1998ta} the Goldstone mode analysis of the D3-brane, the M-theory membrane and fivebrane was carried out and used to determine the effective equations of motion for each of those objects. In each case the analysis gave the description of those objects in a flat background. Some further thought shows that this is the correct answer. The Goldstone mode analysis must give the equations of motion of the string in a flat background since the solution for which one is determining the moduli is that of string in a flat background. 

A string solution in the background of other strings i.e. a string $\sigma$-model in a string background would be a different solution and as such obey a different set of equations of motion. Describing this more technically, to find the $\sigma$-model in a nontrivial background one must find the backreacted wave solution not for asympotically flat space but for one with asymptotically switched on NS-fluxes and then determine its moduli and their equations of motion. Of course, how we normally proceed with brane actions is that once one has determined the effective equations of motion through a Goldstone mode analysis one then covariantizes these equations (in terms of the geometry of moduli space) to determine the general equations of motion. In terms of the doubled string above, this would imply just replacing the flat target space generalized metric with the generalized metric of an arbitrary background. (The quantum properties of such twisted chiral bosons with an arbitrary target space may well be very nontrivial, an analysis of such is outside the scope of the current paper.)

\section{The Membrane as a Wave}
\label{sec:M2brane}
In a similar manner to the string, the membrane will be shown to arise from a massless solution corresponding to a wave in an extended geometry. We will demonstrate this for the membrane in the $SL(5)$ duality invariant theory though it is imagined that this will be true of all the extended geometries corresponding to the exceptional groups. We begin with the equations of motion of the $SL(5)$ theory. The actions of the U-duality manifest theories have been explored at length \cite{Berman:2011jh} but the equations of motion will require the construction of projectors just as in the $O(d,d)$ case since we should only consider variations of the actions that preserve the generalized metric coset structure. We begin by describing these projectors.

\subsection{The $SL(5)$ Duality Invariant Theory}
\label{sec:SL5eoms}
Let us start by examining the extended geometry of the $SL(5)$ duality invariant theory. This arises from the full eleven-dimensional theory by splitting the dimensions into 4+7. The U-duality group acts on the four dimensions and can be made manifest by including the six dual dimensions corresponding to membrane wrappings. There is then a (4+6)-dimensional extended space with manifest $SL(5)$ invariance and no dependence on the remaining seven dimensions. Referring to the $E_{11}$ decomposition into $SL(5)\times GL(7)$, schematically a generalized metric of such an (10+7)-dimensional space can be written as (see \cite{Malek:2012pw})
\begin{equation}
\HH = {\det g_{11}}^{-1/2}
\begin{pmatrix}
\tMM & 0 \\ 0 & g_7
\end{pmatrix}
\end{equation} 
where $\tMM$ is the generalized metric on the extended space and $g_7$ is the metric on the remaining seven dimensions. The conformal factor up front is important as it relates these two otherwise independent sectors, it is given in terms of the determinant of $g_{11}$, the metric of the full eleven-dimensional space.

This $\tMM_{MN}$ is the generalized metric as first given in \cite{Berman:2010is}. It parametrizes the coset $SL(5)/SO(5)$ in terms of the spacetime metric $g_{\mu\nu}$ and the form field $C_{\mu\nu\rho}$
\begin{equation}
\tMM_{MN} = 
\begin{pmatrix}
g_{\mu\nu} + \frac{1}{2}C_{\mu\rho\sigma}g^{\rho\sigma,\lambda\tau}C_{\lambda\tau\nu} 
		& \frac{1}{\sqrt{2}}C_{\mu\rho\sigma}g^{\rho\sigma,\lambda\tau} \\
\frac{1}{\sqrt{2}}g^{\rho\sigma,\lambda\tau}C_{\lambda\tau\nu} & g^{\rho\sigma,\lambda\tau}
\end{pmatrix}
\label{eq:SL5metric}
\end{equation}
for coordinates $X^M = (x^\mu,y_{\mu\nu})$ in the $\mathbf{10}$ of $SL(5)$ and with $g^{\mu\nu,\rho\sigma}=\frac{1}{2}(g^{\mu\rho}g^{\nu\sigma}-g^{\mu\sigma}g^{\nu\rho})$ which is used to raise an antisymmetric pair of indices. Note that there is no overall factor in front, this metric has a determinant of $g^{-2}$ where $g$ is the determinant of the four-metric $g_{\mu\nu}$. Therefore in this form it is actually an element of $GL(5)$, not $SL(5)$. This can be remidied by considering the following.

The theory contains a scaling symmetry for the $GL(5)$ which can be used to rescale $\tMM_{MN}$ by $g$, e.g. $\MM_{MN} = g^{1/5}\tMM_{MN}$ (this particular rescaling leads to a generalized metric with unit determinant, i.e. $\det \MM_{MN}=1$). Noting that $\det g_{11} = g\det g_7$ and assuming a simple form\footnote{For example when considering the compactification of the seven dimensions on a seven-torus with equal radius $R$ this is just $g_7=R\delta_7$ and thus $V=R^7$.} for the seven-metric such that $\det g_7 = V$ we have
\begin{equation}
\HH = 
\begin{pmatrix}
V^{-1/2}g^{-1/2}g^{-1/5}\MM & 0 \\ 0 & V^{-5/14}g^{-1/2}\delta_7
\end{pmatrix} =
\begin{pmatrix}
e^{-\Delta}\MM & 0 \\ 0 & e^{-5\Delta/7}\delta_7
\end{pmatrix} \, .
\end{equation} 
Under an $SL(5)$ transformation the seven-sector should remain unchanged, therefore we have the following $SL(5)$ scalar density
\begin{equation}
e^\Delta = V^{1/2}g^{7/10}
\end{equation}
which we will us to write down the correctly weighted action for the extended theory. In terms of the generalized metric $\MM_{MN}$ with unit determinant and the volume factor $\Delta$ the action reads
\begin{equation}
S = \int \dd^D X e^\Delta R
\label{eq:SL5action}
\end{equation}
where the scalar $R$ is given by
\begin{equation}
\begin{aligned}
R &= \frac{1}{12}\MM^{MN}\partial_M\MM^{KL}\partial_N\MM_{KL}
		-\frac{1}{2}\MM^{MN}\partial_M\MM^{KL}\partial_L\MM_{KN} \\
	&\qquad  + \partial_M\MM^{MN}\partial_N\Delta 
		+ \frac{1}{7}\MM^{MN}\partial_M\Delta\partial_N\Delta \, .
\end{aligned}
\label{eq:SL5R}
\end{equation}
The first two terms reproduce the Einstein-Hilbert and Maxwell term upon imposing section condition. The last two terms are kinetic terms for $\Delta$. The equations of motion for $\Delta$ can be found by varying the action and are given up to total derivatives by $R=0$.

On the other hand, varying the action with respect to the generalized metric and integrating by parts gives
\begin{equation}
\begin{aligned}
\delta S = \int \dd^D X e^{\Delta} &\left[\frac{1}{12}\left(\partial_M \MM^{KL}\partial_N \MM_{KL}
		- 2 \partial_K \MM^{KL}\partial_L \MM_{MN}  
		- 2 \MM^{KL} \partial_K \partial_L \MM_{MN} \right.\right.\\
	&\quad\left.\left.
		+ 2 \MM^{KL}\MM^{PQ}\partial_K \MM_{MP} \partial_L \MM_{NQ} 
		- 2 \MM^{KL}\partial_K\Delta \partial_L \MM_{MN}\right) \right. \\
	&\quad\left. -\frac{1}{2}\left(\partial_M \MM^{KL}\partial_L \MM_{KN}
		- 2 \partial_L \MM^{KL}\partial_M \MM_{KN} 
		- 2 \MM^{KL} \partial_L\partial_M \MM_{KN} \right.\right. \\
	&\quad\left.\left.
		+ 2 \MM^{KP}\MM^{LQ}\partial_{(K} \MM_{M)Q} \partial_L \MM_{NP} 
		- 2 \MM^{KL} \partial_K \Delta \partial_M \MM_{LN}\right) \right.\\
	&\quad\left. -\partial_M\partial_N\Delta - \frac{6}{7} \partial_M\Delta\partial_N\Delta	
		\right] \delta \MM^{MN} \, . 
\end{aligned}
\end{equation}

Note that there is no term for varying $e^{\Delta}$. This factor contains information about the determinant of $\MM_{MN}$ but does not change if the metric is varied as it is fixed to have unit determinant. We will denote everything inside the brackets by $K_{MN}$
\begin{equation}
\delta S = \int \dd^DX e^\Delta K_{MN} \delta \MM^{MN} \, .
\label{eq:SL5varaction}
\end{equation}

As in the case of DFT, \eqref{eq:SL5varaction} does not have to vanish for any variation $\delta \MM^{MN}$ since the generalized metric is constrained to parametrize a coset space. This gives rise to a projector to eliminate the additional degrees of freedom. To impose this constraint and find this projector, one has to use the chain rule. In order to vary the generalized metric with respect to the spacetime metric and the C-field, it will be usefull to use indices $a = \{\mu,5\}$ in the $\mathbf{5}$ of $SL(5)$. The coordinates are then
\begin{equation}
X^M = X^{ab} = 
\begin{cases}
X^{\mu 5} &= x^\mu \\
X^{\mu\nu} &= \frac{1}{2}\epsilon^{\mu\nu\rho\sigma}y_{\rho\sigma}
\end{cases}
\end{equation}
where $\epsilon^{\mu\nu\rho\sigma}$ is the permutation symbol in four dimensions, a tensor density. The generalized metric and its inverse take the form
\begin{equation}
\MM_{ab,cd} =
\begin{pmatrix}
\MM_{\mu 5,\nu 5} & \MM_{\mu 5,\lambda\tau} \\ \MM_{\rho\sigma,\nu 5} & \MM_{\rho\sigma,\lambda\tau}
\end{pmatrix}
= g^{1/5}
\begin{pmatrix}
g_{\mu\nu} + \frac{1}{2}C_{\mu\rho\sigma}g^{\rho\sigma,\lambda\tau}C_{\lambda\tau\nu}
	 & -\frac{1}{2\sqrt{2}} C_{\mu\rho\sigma} g^{\rho\sigma,\alpha\beta} 
	 		\epsilon_{\alpha\beta\lambda\tau} \\
	-\frac{1}{2\sqrt{2}} \epsilon_{\rho\sigma\alpha\beta}
			g^{\alpha\beta,\lambda\tau}C_{\lambda\tau\nu}
	 & g^{-1}g_{\rho\sigma,\lambda\tau}
\end{pmatrix} \notag
\end{equation}
\begin{equation}
\MM^{ab,cd} = g^{-1/5}
\begin{pmatrix}
g^{\mu\nu}  & \frac{1}{2\sqrt{2}}g^{\mu\nu}C_{\nu\alpha\beta}\epsilon^{\alpha\beta\lambda\tau} \\
		\frac{1}{2\sqrt{2}}\epsilon^{\rho\sigma\alpha\beta}C_{\alpha\beta\mu}g^{\mu\nu} 
	& gg^{\rho\sigma,\lambda\tau} + \frac{1}{8}\epsilon^{\rho\sigma\alpha\beta}C_{\alpha\beta\mu} 
		g^{\mu\nu}C_{\nu\gamma\delta}\epsilon^{\gamma\delta\lambda\tau}
\end{pmatrix}
\end{equation}
with $g^{\mu\nu,\alpha\beta}g_{\alpha\beta,\rho\sigma} = \frac{1}{2}(\delta^\mu_\rho\delta^\nu_\sigma - \delta^\mu_\sigma\delta^\nu_\rho)$. Note the factor of $g^{1/5}$ up front since this is the rescaled metric with unit determinant. Using the chain rule and varying the metric in \eqref{eq:SL5varaction} with respect to $\delta g_{\mu\nu}$ and $\delta C_{\mu\nu\rho}$ gives
\begin{align}
\delta S &= \int \dd^DX K_{MN} 
	\left[\frac{\delta \MM^{MN}}{\delta g_{\mu\nu}}\delta g_{\mu\nu} 
		+ \frac{\delta \MM^{MN}}{\delta C_{\mu\nu\rho}}\delta C_{\mu\nu\rho}\right] \\
	&= \int \dd^DX g^{-1/5}\left\{
	\left[-K_{\alpha 5,\beta 5} g^{\alpha(\mu}g^{\nu)\beta} 
		-2 K_{\alpha 5,\beta\beta'}\frac{1}{2\sqrt{2}}g^{\alpha(\mu}g^{\nu)\alpha'}
			C_{\alpha'\gamma\gamma'}\epsilon^{\gamma\gamma'\beta\beta'}\right.\right. \notag\\
	&\left.\left.\hspace{3.2cm} 
		+ K_{\alpha\alpha',\beta\beta'}\left(\vphantom{\frac{1}{8}}
		gg^{\mu\nu}g^{\alpha\alpha',\beta\beta'} 
		- gg^{\alpha(\mu}g^{\nu)[\beta}g^{\beta']\alpha'}
		- gg^{\alpha[\beta}g^{\beta'](\mu}g^{\nu)\alpha'} \right.\right.\right. \notag\\
	&\left.\left.\left.\hspace{3.2cm}- \frac{1}{8}\epsilon^{\alpha\alpha'\gamma\gamma'}
			C_{\gamma\gamma'\sigma}g^{\sigma(\mu}g^{\nu)\sigma'}
			C_{\sigma'\lambda\lambda'}\epsilon^{\lambda\lambda'\beta\beta'}\right) - \frac{1}{5}g^{1/5}K_{MN}\MM^{MN}g^{\mu\nu}
		\right]\delta g_{\mu\nu}\right. \notag\\
	&\left. \hspace{2.7cm} + \left[
		2K_{\alpha 5,\beta\beta'}\frac{1}{2\sqrt{2}}g^{\alpha\alpha'}
			\delta_\gamma^{[\mu}\delta_{\gamma'}^\nu\delta_\sigma^{\rho]}
			\epsilon^{\gamma\gamma'\beta\beta'} \right.\right.\notag\\
	&\left.\left. \hspace{3.2cm} + 2K_{\alpha\alpha',\beta\beta'}
			\frac{1}{8}\epsilon^{\alpha\alpha'\gamma\gamma'}
			\delta_\gamma^{[\mu}\delta_{\gamma'}^\nu\delta_\sigma^{\rho]}
			g^{\sigma\sigma'}C_{\sigma'\lambda\lambda'}\epsilon^{\lambda\lambda'\beta\beta'}
			\right]\delta C_{\mu\nu\rho}	\right\}
\end{align}
where the term $\frac{1}{5}K_{MN}\MM^{MN}g^{\mu\nu}\delta g_{\mu\nu}$ arises from varying the determinant factor. After cleaning up and dropping the symmetrizing and antisymmetrizing brackets, the $g$'s and $C$'s are re-expressed in terms of $\MM$ (factors of $g^{1/5}$ have to be accounted for carefully)
\begin{equation}
\begin{aligned}
\delta S &= \int \dd^DX \left\{g^{1/5}\vphantom{\frac{1}{\sqrt{2}}}
	\left[-K_{\alpha 5,\beta 5} \MM^{\alpha 5,\mu 5}\MM^{\nu 5,\beta 5}
		-2 K_{\alpha 5,\beta\beta'}\MM^{\alpha 5,\mu 5}\MM^{\nu 5,\beta\beta'} 
			\right.\right. \\
	&\left.\left. \hspace{2.5cm}
		+ K_{\alpha\alpha',\beta\beta'}\left(g^{-1/5}\MM^{\mu 5,\nu 5}gg^{\alpha\alpha',\beta\beta'}
			- \MM^{\alpha\alpha',\mu 5}\MM^{\nu 5,\beta\beta'}\right)  \right.\right. \\
	&\left.\left.\hspace{2.5cm} - \frac{1}{5} K_{MN}\MM^{MN}\MM^{\mu 5,\nu 5}\right]\delta g_{\mu\nu}\right. \\
	&\left.\hspace{2cm} +\frac{1}{\sqrt{2}}\left[
		  K_{\alpha 5,\beta\beta'} \MM^{\alpha 5,\mu 5}\epsilon^{\nu\rho\beta\beta'}
		+ K_{\alpha\alpha',\beta\beta'} \MM^{\alpha\alpha',\mu 5}
			\epsilon^{\nu\rho\beta\beta'}\right]\delta C_{\mu\nu\rho}	\right\}
\end{aligned}
\end{equation}
Now the indices can be re-covariantized to be expressed as
\begin{equation}
\begin{aligned}
\delta S &= \int \dd^DX \left\{\vphantom{\frac{1}{\sqrt{2}}}
		g^{1/5}K_{KL}\left(\MM^{M, \mu 5}\MM^{\nu 5, N}\MM_{MP}	
			\frac{1}{4}\epsilon^{aPK}\epsilon_{aNQ}\MM^{QL} 
		- \MM^{K, \mu 5}\MM^{\nu 5, L}  \right.\right. \\
	&\left.\left.\hspace{4cm}	- \frac{1}{5}\MM^{KL}\MM^{\mu 5,\nu 5}\right)\delta g_{\mu\nu} 
	 	+ \frac{1}{\sqrt{2}}K_{KL}\MM^{K, \mu 5}\epsilon^{\nu\rho L5}\delta C_{\mu\nu\rho}
	 	\right\}
\end{aligned}
\label{eq:projderivation}
\end{equation}
which reproduces the previous line if the extended indices are expanded and summed over. In a final step these expressions can be written in terms of a projected set of equations
\begin{equation}
\delta S =\int \dd^DX (-3) \PP{M}{N}{K}{L}K_{KL}
	\left(g^{1/5}\MM^{M, \mu 5}\MM^{\nu 5, N}\delta g_{\mu\nu}
	 - \frac{1}{2\sqrt{2}}\MM^{M, \mu 5}\epsilon^{\nu\rho N5}\delta C_{\mu\nu\rho}\right)
\end{equation}
where the projector is given by
\begin{equation}
\PP{M}{N}{K}{L} = \frac{1}{3}\left({\delta_M}^{(K}{\delta_N}^{L)}
		+ \frac{1}{5}\MM_{MN}\MM^{KL}
		- \frac{1}{4}\MM_{MP}\epsilon^{aP(K}\epsilon_{aNQ}\MM^{L)Q} \right)
\end{equation}
which is symmetric in both $MN$ and $KL$ as can be seen from the contraction with the symmetric $\delta g_{\mu\nu}$ and $K_{KL}$ respectively. Note that the term containing $\delta C_{\mu\nu\rho}$ does not impose any symmetry property on the projector.

The variation of the action has to vanish for \emph{any} $\delta g_{\mu\nu}$ and $\delta C_{\mu\nu\rho}$ independently, therefore the equations of motion are given by  
\begin{equation}
\PP{M}{N}{K}{L}K_{KL} = 0
\label{eq:SL5eom}
\end{equation}
with $K_{MN}$ defined in \eqref{eq:SL5varaction}.

\subsection{Divertimento: Equations of Motion with a Projector}
\label{sec:divertimento}
In general, the dynamics of extended geometry can be described using a projected equation of motion. The action is given by
\begin{equation}
S = \int \dd^{D}X \LL
\end{equation}
where the Lagrangian $\LL$ includes the integration measure for the extended space. Setting the variation of the action to zero gives
\begin{equation}
\delta S = \int \dd^{D}X K_{MN}\delta \MM^{MN} = 0 
\end{equation}
where $K_{MN}=\delta\LL/\delta\MM^{MN}$ is the variation of the Lagrangian with respect to the generalized metric. The integrand does not have to vanish for any $\delta\MM^{MN}$ since the generalized metric is constraint to parametrize the coset space $G/H$. This constraint gives rise to a projector in the equations of motion
\begin{equation}
{P_{MN}}^{KL}K_{KL} = 0.
\label{eq:genEoM}
\end{equation}
The extended geometries are all equipped with the so called $Y$-tensor described in \cite{Berman:2012vc}. The $Y$-tensor determines the deviation from usual geometry in that it gives the correction to the Lie derivative to form the generalized Lie derivative given in \cite{Berman:2011cg}. Following the method for $O(d,d)$ and then $SL(5)$ where we use a chain rule type arguement, we see that the projector may be written in a standard form using only the generalized metric and the $Y$-tensor
\begin{equation}
{P_{MN}}^{KL} = \frac{1}{a}\left( {\delta_M}^{(K}{\delta_N}^{L)}
		+ b \MM_{MN}\MM^{KL}	- \MM_{MP}{Y^{P(K}}_{NQ}\MM^{L)Q} \right) \, ,
\label{eq:genProj}
\end{equation}
together with the constants $a$ and $b$ which depend on the dimension of the extended space $D$ and thus the U-duality group. These constants together with the $Y$-tensor are given in the following table for some of the duality groups under consideration.
\begin{equation}
\begin{array}{|l|c|ccc|}
\hline
        & {Y^{MN}}_{KL} 								& a  	& b	& D 		\\ \hline
O(d,d)  & \eta^{MN}\eta_{KL}    						& 2     		& 0 		& 2d 	\\ 
SL(5)   & \frac{1}{4}\epsilon^{iMN}\epsilon_{iKL}   	& 3  	 	& 1/5  	& 10    	\\ 
SO(5,5) & \frac{1}{2}(\Gamma^i)^{MN}(\Gamma_i)_{KL}  & 4 			& 1/4   	& 16		\\ \hline
\end{array}
\label{eq:ytensor}
\end{equation}
The elements that form the $Y$-tensor are $\eta_{MN}$, the invariant metric of $O(d,d)$; $\epsilon_{iMN}=\epsilon_{iabcd}$, the $SL(5)$ alternating tensor ($i=1,\dots,5$); and ${(\Gamma^i)^M}_N$, the $16\times 16$ Majorana-Weyl representation of the $SO(5,5)$ Clifford algebra ($i=1,\dots,10$).

Our ${P_{MN}}^{KL}$ is a genuine projector in the sense that $P^2=P$ and its eigenvalues are either $0$ or $1$. The eigenvectors with eigenvalue $0$ span the kernel of the projector. Those parts of $K_{MN}$ proportional to these eigenvectors are projected out and eliminated from the equations of motion. 

The multiplicity of the eigenvalues $0$ and $1$ are called nullity (dimension of the kernel) and rank of the projector respectively. They add up to the dimension $D$ of the vector space of eigenvectors. We have not shown that this is true beyond the groups in the table above since the calculations have been done just by brute force. However,  given the structure of the exceptional geometric theories, in that the theories up to $E_7$ are completely determined by the generalized metric and the $Y$-tensor (along with a few dimensionally dependent constants), then we expect this projector to be true at least up to $E_7$ with only the constants $a$ and $b$ to be determined.

Note, the object $K_{MN}$ is symmetric and thus has $\frac{1}{2}D(D+1)$ independent components in a generalized space with $D$ dimensions. The bosonic degrees of freedom of the theories under consideration are given by the metric tensor $g_{\mu\nu}$ and the form fields $B_{\mu\nu}$ or $C_{\mu\nu\rho}$ (plus one for the dilaton $\phi$ in DFT and the volume factor $\Delta$ in the $SL(5)$ theory). One equation of motion is needed for each of those degrees of freedom. The projector reduces the components of the equation $K_{MN}=0$ such that the right number of independent equations remain.

\subsection{Wave Solution or Membrane in the $SL(5)$ Theory}
\label{sec:SL5wave}
The wave solution for the $SL(5)$ duality invariant theory is given by a generalized metric $\MM_{MN}$ with line element
\begin{equation}
\begin{aligned}
\dd s^2 &= \MM_{MN}\dd X^M \dd X^N \\
		&= (H-2)\left[(\dd x^1)^2 - (\dd x^2)^2 - (\dd x^3)^2 \right] 
			+ (\dd x^4)^2 \\
		&\quad + 2(H-1)\left[\dd x^1\dd y_{23} 
			+ \dd x^2\dd y_{13} - \dd x^3\dd y_{12} \right] \\
		&\quad	- H\left[(\dd y_{13})^2 + (\dd y_{12})^2 - (\dd y_{23})^2 \right]
		 	+ (\dd y_{34})^2  + (\dd y_{24})^2 - (\dd y_{14})^2.
\end{aligned}
\label{eq:SL5ppwave}
\end{equation}
This generalized metric solves the equations of motion of the $SL(5)$ theory derived in \ref{sec:SL5eoms} (see Appendix \ref{sec:SL5check}). It can be interpreted as a pp-wave in the extended geometry which carries momentum in the directions dual to $x^2$ and $x^3$ i.e. combinations of $y_{12}, y_{13}$ and $y_{23}$. Since it is a pp-wave it has no mass or charge and the solution is pure metric, there is no form field it couples to. As before, $H$ is a harmonic function of the transverse coordinate $x^4$: $H=1+h\ln x^4$. It is smeared in the remaining dual directions. 

A Kaluza-Klein ansatz suitable for the geometry here that allows us to rewrite the solution in terms of four-dimensional quatities and reducing the dual directions is
\begin{equation}
\begin{aligned}
\dd s^2 	&= \left(g_{\mu\nu} + e^{2\phi}C_{\mu\lambda\tau}g^{\lambda\tau,\rho\sigma}
			C_{\rho\sigma\nu}\right)\dd x^\mu \dd x^\nu \\
	&\quad + 2e^{2\phi}C_{\mu\lambda\tau}g^{\lambda\tau,\rho\sigma}
		\dd x^\mu \dd y_{\rho\sigma} 
	+ e^{2\phi}g^{\lambda\tau,\rho\sigma}\dd y_{\lambda\tau}\dd y_{\rho\sigma}.
\end{aligned}
\label{eq:KKforSL5}
\end{equation}
The factor $e^{2\phi}$ is a scale factor and needs to be included for consistency. This decompostion of the generalized metric into the usual metric and C-field  resembles the form of the generalized metric \eqref{eq:SL5metric} as in the DFT case.

By comparing \eqref{eq:KKforSL5} with \eqref{eq:SL5ppwave}, the fields of the reduced system with coordinates $x^\mu$ can be computed. From the diagonal terms we find
\begin{equation}
g_{\mu\nu} = \diag (-H^{-1}, H^{-1}, H^{-1}, 1) \qandq 
g^{\mu\nu,\rho\sigma} = e^{-2\phi}\diag (-H, -H, -1, H, 1, 1)
\end{equation}
and since $g^{\mu\nu,\rho\sigma}$ is given by $g^{\mu\nu}$, the inverse of $g_{\mu\nu}$, we need $e^{2\phi}=H^{-1}$ for consistency. The corresponding line element is
\begin{equation}
\dd s^2 = -H^{-1}\left[(\dd x^1)^2 - (\dd x^2)^2 - (\dd x^3)^2 \right] + (\dd x^4)^2.
\label{eq:membraneKK}
\end{equation}
The off-diagonal terms give the antisymmetric C-field whose only non-zero component is 
\begin{equation}
C_{123} = -(H^{-1}-1).
\end{equation}
This metric and C-field look like the membrane in M-theory. To complete this identification, \eqref{eq:membraneKK} has to be rescaled to be expressed in the Einstein frame.

The standard rescaling procedure (in four dimensions) gives
\begin{equation}
g_{\mu\nu} = \Omega^{-2}\tg_{\mu\nu} = H^{-3/2}\tg_{\mu\nu}
\end{equation}
where 
\begin{equation}
\Omega^2 = \sqrt{|\det e^{2\phi} g^{\mu\nu,\rho\sigma}|} = H^{3/2}.
\end{equation}
Therefore the rescaled metric reads $\tg_{\mu\nu}=H^{3/2}g_{\mu\nu}$ and the full solution in the Einstein frame is\footnote{The C-field is unaffected by the rescaling, only its field strength obtains a different factor in the action.}
\begin{equation}
\dd s^2 = -H^{-1/2}\left[(\dd x^1)^2-(\dd x^2)^2-(\dd x^3)^2 \right]+H^{3/2}(\dd x^4)^2 
\label{eq:membrane}
\end{equation}
which is indeed the M2-brane in four dimensions in the Einstein frame. The membrane is extended in the $x^2-x^3$ plane. We have thus shown that the solution \eqref{eq:SL5ppwave} which carries momentum in the directions dual to $x^2$ and $x^3$ in the extended geometry corresponds to a membrane stretched along these directions from a reduced point of view. By similar arguments as in the string case, the mass and charge of the M2-brane are given by the momenta in the dual directions.

\subsection{Goldstone Modes of the Wave Solution}
\label{sec:SL5goldstones}
Following the same procedure as for the DFT wave we will now perform the Goldstone mode analysis for the wave in $SL(5)$. To do this we will use the five-dimensional coordinate representation introduced above and split the coordinates into worldvolume and transverse parts. Note that the membrane in four dimensions only has one transverse direction. By introducing $m,n=1,2,3$, the coordinates read
\begin{equation}
X^M = X^{ab} = (X^{m5};X^{45},X^{m4},X^{mn}) = (x^m;x^4,y^{mn},y^{m4}).
\end{equation}
In this notation the non-zero components of the generalized metric for the $SL(5)$ wave given in \ref{eq:SL5ppwave} can be written as
\begin{equation}
\begin{aligned}
\MM_{m5,n5} &= (2-H)\II_{mn}			&	\MM^{m5,n5} &= H\II^{mn} 		\\
\MM_{m4,n4} &= -H\II_{mn}			&	\MM^{m4,n4} &= -(2-H)\II^{mn} 	\\
\MM_{m4,n5} &= -(H-1)\II_{mn}		&	\MM^{m4,n5} &= -(H-1)\II^{mn} 	\\
\MM_{mn,kl} &= \II_{mn,kl}			&	\MM^{mn,kl} &= \II^{mn,kl}		\\
\MM_{45,45} &= 1						&	\MM^{45,45} &= 1
\end{aligned}
\end{equation}
where the harmonic function $H$ is a function of $X^{45}=x^4$ only and for convenience these two matrices are introduced
\begin{align}
\II_{mn} &=  \begin{pmatrix} -1 & 0 & 0 \\ 0 & 1 & 0 \\ 0 & 0 & 1 \end{pmatrix} = \II^{mn} \, ,  &
\II_{mn,kl} &= \begin{pmatrix} 1 & 0 & 0 \\ 0 & 1 & 0 \\ 0 & 0 & -1 \end{pmatrix} = \II^{mn,kl}.
\end{align}
The generalized Lie derivative of the metric and the volume factor (a density) are given by the same expressions as before (cf. \eqref{eq:genLieMetric} and \eqref{eq:genLieDilaton}) with the $Y$-tensor for $SL(5)$ being
\begin{equation}
{Y^{MN}}_{KL} = \frac{1}{4}\epsilon^{aMN}\epsilon_{aKL}
\end{equation}
where these are five-dimensional permutation symbols which are tensor densities. We thus have
\begin{align}
\LL_\xi\MM_{MN} &= \xi^L\partial_L\MM_{MN} + 2\MM_{L(M}\partial_{N)}\xi^L 
	- \frac{1}{2}\MM_{L(M|}\epsilon_{a|N)Q}\epsilon^{aLP}\partial_P\xi^Q \\
\LL_\xi\Delta &= \xi^M\partial_M\Delta + \partial_M\xi^M.
\end{align}

We again pick a transformation parameter $\xi^M$ with non-zero components only in the transverse directions but not along the worldvolume (and its dual). This can be described by an $SL(5)$ vector field
\begin{equation}
\xi^M=\xi^{ab}=(0,H^\alpha\hphi,0,H^\beta\htphi^{mn})
\end{equation}
where $\hphi$ and $\htphi^{ij}$ are a constant scalar and dualized vector that later will become the Goldstone modes once they are allowed a dependence on the worldvolume coordinates, $H$ is the harmonic function and $\alpha, \beta$ are constants determined by normalisability. 

Using the generalized Lie derivative given above we compute $m_{MN} = \LL_\xi\MM_{MN}$ 
\begin{equation}
\begin{aligned}
m_{m5,n5} &= m_{m4,n4} = m_{m4,n5} = -\II_{mn}\hphi H^\alpha \partial H \\
m_{45,45} &= 2\hphi \partial H^\alpha \\
m_{mn,kl} &= -\II_{mn,kl}\hphi \partial H^\alpha \\
m_{mn,45} &= \frac{1}{2}\II_{mn,kl} \htphi^{kl} \partial H^\beta
\end{aligned}
\end{equation}
and, recalling that $\Delta$ is a constant for our solution,
\begin{equation}
\lambda = \LL_\xi\Delta = \partial(\hphi H^\alpha).
\end{equation}
Now the four modes $\phi, \tphi^{12}, \tphi^{13}$ and $\tphi^{23}$ are allowed to depend on the worldvolume coordinates $x^m$ (and the hats are removed). 

For the equations of motion we need $K_{MN}$ and $R$ only with terms with two derivatives on $m_{MN}$ and $\lambda$. There are no such terms in $R$ as given in \eqref{eq:SL5R} but upon integrating by parts they can arise. We thus have
\begin{align}
K_{MN} &= \MM^{KL}\partial_K\partial_{(M}m_{N)L} 
	- \frac{1}{6}\MM^{KL}\partial_K\partial_L m_{MN}
	- \partial_M\partial_N \lambda \\
R &= -\frac{2}{7}\MM^{MN}\partial_M\partial_N\lambda - \partial_M\partial_N m^{MN}
\end{align}
Inserting $m_{MN}$ and defining $\Box\phi=H\II^{mn}\partial_m\partial_n\phi$ this gives 
\begin{equation}
\begin{aligned}
K_{m5,n5} &= -(1+\alpha H^{-1})\partial_m\partial_n\phi (H^\alpha\partial H) 
	+ \frac{1}{6}\II_{mn}\Box\phi (H^\alpha\partial H)  \\
K_{m4,n4} &= \frac{1}{6}\II_{mn}\Box\phi (H^\alpha\partial H) \\
K_{m5,n4} &= -\frac{1}{2}\partial_m\partial_n\phi (H^\alpha\partial H) 
	+ \frac{1}{6}\II_{mn}\Box\phi (H^\alpha\partial H) \\
K_{45,45} &= -\frac{\alpha}{3}H^{-1} \Box\phi (H^\alpha\partial H) \\
K_{mn,kl} &= \frac{\alpha}{6}H^{-1}\II_{mn,kl} \Box\phi (H^\alpha\partial H) \\
K_{mn,45} &= -\frac{\beta}{12}H^{-1}\II_{mn,kl} \Box\tphi^{kl} (H^\beta\partial H) \\
K_{m5,45} &= -\frac{1}{2}\partial_m\phi \partial(H^\alpha\partial H) \\
K_{m4,45} &= -\frac{1}{2}\partial_m\phi \partial(H^\alpha\partial H)\\
K_{m5,kl} &= \frac{1}{4}\II_{kl,pq}\partial_m\tphi^{pq}\partial^2H^\beta \\
K_{m4,kl} &= 0
\end{aligned}
\end{equation}
The volume factor equation gives
\begin{equation}
R = H^{-1}(\frac{\alpha}{7}+1)\Box\phi(H^\alpha\partial H) = 0
\end{equation}
which is solved by $\Box\phi = 0$.

Now we have 14 components of the projected equation of motion ${P_{MN}}^{KL}K_{KL}=0$:
\begin{itemize}

\item three of the form $K_{m5,45} \sim K_{kl,n4} + K_{kl,n5}$
\begin{equation}
\begin{aligned}
K_{15,45} &= (H-2)(K_{12,24} + K_{13,34}) - (H-1)(K_{12,25} + K_{13,35}) \\
K_{25,45} &= (H-2)(K_{12,14} + K_{23,34}) - (H-1)(K_{12,15} + K_{23,35}) \\
K_{35,45} &= (H-2)(K_{13,14} - K_{23,24}) - (H-1)(K_{13,15} - K_{23,25})
\end{aligned}
\end{equation}

\item three of the form $K_{m4,45} \sim K_{kl,n4} + K_{kl,n5}$
\begin{equation}
\begin{aligned}
K_{14,45} &= (H-1)(K_{12,24} + K_{13,34}) - H(K_{12,25} + K_{13,35}) \\
K_{24,45} &= (H-1)(K_{12,14} + K_{23,34}) - H(K_{12,15} + K_{23,35}) \\
K_{34,45} &= (H-1)(K_{13,14} - K_{23,24}) - H(K_{13,15} - K_{23,25})
\end{aligned}
\end{equation}

\item three of the form $K_{mn,kl} \sim K_{p4,q4} + K_{p4,q5} + K_{p5,q4} + K_{p5,q5}$ with $mn\neq kl$
\begin{equation}
\begin{aligned}
K_{13,23} &= (H-2)K_{14,24} - (H-1)K_{14,25} - (H-1)K_{15,24} - H K_{15,25} \\
-K_{12,23} &= (H-2)K_{14,34} - (H-1)K_{14,35} - (H-1)K_{15,34} - H K_{15,35} \\
-K_{12,13} &= (H-2)K_{24,34} - (H-1)K_{24,35} - (H-1)K_{25,34} - H K_{25,35}
\end{aligned}
\end{equation}

\item two relating the $K_{m4,m4}$, $K_{m4,m5}$ and $K_{m5,m5}$ components
\begin{equation}
\begin{aligned}
H(K_{15,15}-K_{25,25}-K_{35,35}) &=  (H-2)(K_{14,14}-K_{24,24}-K_{34,34}) \\
H(K_{14,15}-K_{24,25}-K_{34,35}) &=  (H-1)(K_{14,14}-K_{24,24}-K_{34,34}) \\
\end{aligned}
\end{equation}

\item and three relating $K_{mn,kl}$ with $mn=kl$ and $K_{45,45}$ to $K_{m4,m4}$, $K_{m4,m5}$ and $K_{m5,m5}$
\begin{equation}
\begin{aligned}
K_{12,12}-K_{13,13} &= (H-2)(K_{14,14} - 2K_{24,24}) - 2(H-1)(K_{14,15}-2K_{24,25}) \\
		&\qquad + H(K_{15,15}-2K_{25,25}) +\frac{2}{H}(K_{14,14} - K_{24,24} - K_{34,34}) \\
K_{12,12}+K_{23,23} &= (H-2)(2K_{14,14} - K_{24,24}) - 2(H-1)(2K_{14,15} - K_{24,25}) \\
		&\qquad + H(2K_{15,15} - K_{25,25}) +\frac{2}{H}(K_{14,14} - K_{24,24} - K_{34,34})\\
K_{45,45} - 2K_{12,12} &= 2(H-2)(K_{14,14} - K_{24,24}) + 4(H-1)(2K_{14,15} - K_{24,25}) \\
		&\qquad - 2H(K_{15,15} - K_{25,25}) -\frac{3}{H}(K_{14,14} - K_{24,24} - K_{34,34})
\end{aligned}
\end{equation}

\end{itemize}
The first and second block of equations can be combined to get cancellations, resulting in three equations for $\tphi$
\begin{equation}
\begin{aligned}
\partial_2\tphi^{12} + \partial_3\tphi^{13} &= 0 \\
\partial_1\tphi^{12} - \partial_3\tphi^{23} &= 0 \\
\partial_1\tphi^{13} + \partial_2\tphi^{23} &= 0 \, .
\end{aligned}
\end{equation}
Defining $\tphi_i = \frac{1}{2}\epsilon_{ijk}\tphi^{jk}$ this can be written as
\begin{equation}
\begin{aligned}
\partial_2\tphi_3 - \partial_3\tphi_2 &= 0 \\
\partial_1\tphi_3 - \partial_3\tphi_1 &= 0 \\
\partial_2\tphi_1 - \partial_1\tphi_2 &= 0 \, . 
\end{aligned}
\end{equation}

All the remaining blocks of equations are either trivial or satisfied by $\Box\phi=0$. 

One would expect a non-zero right-hand side for the above equations of the form $\partial_m\phi$ to get relations between $\phi$ and $\tphi$ 
\begin{equation}
\partial_m\phi (\partial^2 H^\alpha) \sim -\II_{mn}\epsilon^{npq}\partial_p\tphi_q (\partial^2 H^\beta)
\end{equation}
This does not only provide a condition for $\beta$ to be equal to $\alpha$, but also the three equations needed to reduce the number of modes from four to one.

The reason for the zero on the right-hand side is due to a degeneracy in considering the membrane with its three-dimensional worldvolume in a four-dimensional background. There is only one transverse direction and hence only one contributing derivative $\partial_{45}\equiv\partial$. So a term like 
\begin{equation}
\delta_{mn}\delta^{kl}\partial_k\partial_l H^\alpha - {\delta_m}^k{\delta_n}^l\partial_k\partial_l H^\alpha
\end{equation}
as it arose for the string vanishes for the membrane. It would be interesting to see if the same calculation for the membrane derived from a wave in a larger extended geometry, e.g. the (5+10+1)-dimensional extended space with manifest $SO(5,5)$ invariance along five dimensions, would provide a duality relation between the $\phi$'s and $\tphi$'s that could be turned into a self-duality relation resembling the result in \cite{Duff90b}.

\section{Discussion}

We have seen that strings and branes are null waves from the point of view of extended theories. The BPS nature of these solutions has its origin in the fact that the null wave is BPS and its reduction naturally gives rise to a BPS condition of charge being equal to tension. 

There are immediate natural extensions to this work such as understanding how this works for the supersymmetric theory and checking how this works for other branes such as the M-theory fivebrane. There is also the the more ambitious question as to whether the same analysis works for lower BPS objects such as 1/4 BPS states. 

The Goldstone mode analysis provides a particularly interesting interplay between worldvolume and spacetime approaches. The solutions all obey the section condition and the local symmetry variations used to calculate the Goldstone modes also obey  the section condition but there are still components of the variation, $\tphi$, that are in the extended directions. These are crucial in giving the Tseytlin string. Thus the relation to the section condition in the target space and the chirality condition may be understood as follows. From the point of view of the string world-sheet one should view the $\tphi$ deformations as components of a local symmetry variation in the extended dimensions but one that still does not functionally depend on the dual coordinates. This is crucial since it means the section condition is still obeyed in the Tseytlin string.

Other fascinating possibilities will be to extend this to branes that are non-BPS but are thermodynamically excited. The hope of embedding brane thermodynamics in DFT and extended geometries is intruiging.

From this perspective there is an intriguing possibilitiy of how one should calculate string scattering amplitudes. Many novel contemporary techniques have been developed for understanding the amplitudes of the massless sector of many theories \cite{ArkaniHamed:2012nw}. Now string and branes themselves may be viewed as massless objects all be it in a theory with extra dimensions. These objects being massless degrees of freedom fits well with the idea (previously expressed in \cite{Englert:2003zs}) that one may think of strings and branes as Goldstone modes of the spontaneously broken duality symmetry. As such the appearance of nonlinear realized duality symmetry (see for example \cite{Berman:2011jh} and references therein) in the target space is unsurprising from this perspective. Effective actions of sigma models with nonlinearly realized symmetries in target spaces began with the effective action of pions, the Goldstone modes of broken chiral symmetry.

Another direction of interest is to consider unsmearing the wave solution. It is uncertain whether this can make sense since it will then break the section condition and yet with Scherk-Schwarz theories the section condition is broken and with the localized KK-solution \cite{Jensen:2011jna} the branes become localized in a dual coordinate. Studying the particulars of interesting backgrounds like those described in this paper and their localizations may provide insight into futher possibilities.

\section*{Acknowledgement}
We are very grateful to Chris Blair and Emanuel Malek for concrete and valuable help with the determinant and weight factors in the $SL(5)$ theory. We have also benefited from numerous discussions with Martin Cederwall, Paul Cook, Jeong-Hyuck Park and Malcolm Perry on many aspects of DFT and the extended geometries. DSB and FJR are grateful to the Yukawa Institute for Theoretical Physics in Kyoto for the ``Exotic Structures of Spacetime" meeting where this work was completed. DSB is supported by STFC consolidated grant ST/J000469/1 ``String Theory, Gauge Theory and Duality''. JB and FJR are supported by STFC studentships.

\appendix
\section{Solution Check}
\subsection{The Wave in DFT}
\label{sec:DFTcheck}

In this appendix we proof that the wave solution to DFT presented in \eqref{eq:DFTppwave} does indeed satisfy the equations of motion \eqref{eq:DFTeomDilaton} and \eqref{eq:DFTeom} derived from the DFT action. We will actually show that the stronger equation $K_{MN}=0$ instead of the projected equation is satisfied. 

Both $R$ and $K_{MN}$ have three kind of terms: those just containing the generalized metric $\HH_{MN}$, those containing the dilaton $d$ and those with the generalized vielbein ${\EE^A}_M$. The vielbein terms are always proportional to the Y-tensor and thus vanish since our solution satisfies the section condition. In our solution $d$ is constant so all the dilaton terms vanish as well as they are always acted on by a derivative operator. This leaves us to check the metric terms. Recall that the harmonic function $H$ is a funtion of the transverse $y^m$ only, so the only derivatives acting on $\HH$ that give a non-vanishing contribution are the $\partial_m$.

We will split this task into several steps. First consider the term that is proportional to $\partial_M\HH^{KL}\partial_N\HH_{KL}$. Using the notation of Section \ref{sec:DFTgoldstones}, we can expand the indices to get
\begin{equation}
\begin{aligned}
\partial_M\HH^{KL}\partial_M\HH_{KL} 
	&\rightarrow \partial_m\HH^{ab}\partial_n\HH_{ab}
		+ \partial_m\HH^{kl}\partial_n\HH_{kl} \\
	&\qquad	+ \partial_m\HH^{\ba\bb}\partial_n\HH_{\ba\bb} 
		+ \partial_m\HH^{\bk\bl}\partial_n\HH_{\bk\bl}
		+ 2\partial_m\HH^{a\bb}\partial_n\HH_{a\bb}  \\
	&= \II^{ab}\II_{ab}\partial_m H \partial_n(2-H) 
		+ \II^{\ba\bb}\II_{\ba\bb}\partial_m(2-H)\partial_n H  \\
	&\qquad	+ 2\JJ^{a\bb}\JJ_{a\bb}\partial_m(H-1)\partial_n(H-1) \\	
	&= (-2-2+4)\partial_m H \partial_n H \\
	&= 0. 
\end{aligned}
\label{eq:appendixsum}
\end{equation}

Next consider the term in $R$ proportional to $\partial_M\HH^{KL}\partial_K\HH_{NL}$. It vanishes as well
\begin{equation}
\partial_M\HH^{KL}\partial_K\HH_{NL} \rightarrow \partial_m\HH^{kl}\partial_k\HH_{nl} = 0.
\end{equation} 
Similarly the terms in $K_{MN}$ and $R$ where the derivatives are contracted with the generalized metric (in any combination) vanish since the only derivative we need to consider is $\partial_m$, but upon contraction this forces both indices on $\HH$
to be of $kl$ type and $\HH^{kl}=\delta^{kl}$ so its derivative vanishes.

This leaves us with two more terms in $K_{MN}$ to check
\begin{equation}
\begin{aligned}
\HH^{KL}\partial_K\partial_L\HH_{MN} 
	&\rightarrow \delta^{kl}\partial_k\partial_l\HH_{MN} = 0 \quad\mathrm{since}\quad \delta^{kl}\partial_k\partial_l H=0 \\
\HH^{KL}\HH^{PQ}\partial_K\HH_{MP}\partial_L\HH_{NQ} &\rightarrow \delta^{kl}\HH^{PQ}\partial_k\HH_{MP}\partial_l\HH_{NQ} \, .
\end{aligned}
\end{equation}
The first one vanishes since $\HH_{MN}$ is a linear function of $H$ which is harmonic and thus annihilated by the Laplacian. The second expression has to be expanded for all possible values $MN$ can take. In each case it vanishes either trivially or the terms sum to zero along the lines of \eqref{eq:appendixsum}.

Thus we have shown that all terms in $K_{MN}$and $R$ are zero and therefore the equations of motion is satisfied by our solution.

\subsection{The Wave in $SL(5)$}
\label{sec:SL5check}
As for the DFT solution, we have to check that the solution presented in \eqref{eq:SL5ppwave} is actually a solution to the equations of motion \eqref{eq:SL5eom} and $R=0$ of the extended geometry of $SL(5)$. 

Before we insert the metric into $K_{MN}$ and $R$, we note some simplifications. The harmonic function $H$ is a function only of the transverse coordinates, just $x^4=X^{45}$ in our case. Therefore the only derivative that yields a non-zero result is $\partial_{45}$ which we will simply denote by $\partial$. Thus, just as in the DFT case, terms like
\begin{equation}
\partial_K \MM^{KL}\partial_L \MM_{MN}, \quad
\partial_M \MM^{KL}\partial_L \MM_{KN}, \quad
\MM^{KL} \partial_L\partial_M \MM_{KN},
\end{equation}
that is terms where a derivative acts on a metric which is contracted with a derivative, vanish since $\MM_{45,45}=1$.

The volume factor $\Delta$ is a constant for our solution, so all terms with $\partial \Delta$ also vanish. Furthermore, since $H$ is a harmonic function, it is annihilated by the Laplacian and therefore
\begin{equation}
\MM^{KL} \partial_K \partial_L \MM_{MN} = \partial^2 \MM_{MN} = 0
\end{equation}
since all the components of $\MM_{MN}$ are linear functions of $H$.

With these simplifications in mind, most of the terms in $K_{MN}$ and $R$ vanish trivially. We only need to check two terms explicitly, namely
\begin{equation}
\begin{aligned}
\partial_M \MM^{KL}\partial_N \MM_{KL} \qandq
\MM^{KL} \MM^{PQ} \partial_K \MM_{MP} \partial_L \MM_{NQ}.
\end{aligned}
\label{eq:terms2check}
\end{equation}
Using the notation of Section \ref{sec:SL5goldstones}, we start with the first expression and expand the indices to get
\begin{equation}
\begin{aligned}
\partial_M \MM^{KL}\partial_N \MM_{KL} 
	&\rightarrow \partial\MM^{k5,l5}\partial\MM_{k5,l5}
		+ \partial\MM^{k4,l4}\partial\MM_{k4,l4}
		+ 2\partial\MM^{k5,l4}\partial\MM_{k5,l4}  \\
	&\qquad + \partial\MM^{kl,pq}\partial\MM_{kl,pq}
		+ \partial\MM^{45,45}\partial\MM_{45,45} \\
	&= \II^{mn}\II_{mn}\left[\partial H\partial(2-H) 
		+ \partial(2-H)\partial H
		+ 2\partial(H-1)\partial(H-1) \right]\\
	&= 3\left[-1-1+2\right] \partial H \partial H \\
	&=0.
\end{aligned}
\end{equation}
Similarly we can show that the other expression in \eqref{eq:terms2check} vanishes
\begin{equation}
\begin{aligned}
\MM^{KL}\MM^{PQ}\partial_K\MM_{MP}\partial_L\MM_{NQ} 
	&\rightarrow \MM^{p5,q5}\partial\MM_{M,p5}\partial\MM_{N,q5}
		+ \MM^{p4,q4}\partial\MM_{M,p4}\partial\MM_{N,q4} \\
	&\qquad	+ 2\MM^{p5,q4}\partial\MM_{M,p5}\partial\MM_{N,q4} \\
	&\qquad 	+ \MM^{kl,pq}\partial\MM_{M,kl}\partial\MM_{N,pq} 
		+ \MM^{45,45}\partial\MM_{M,45}\partial\MM_{N,45}\\
	&\rightarrow \left[H-(2-H)-2(H-1)\right]\II^{pq}\II_{mp}\II_{nq}
		\partial H \partial H \\
	&= \left[H-2+H-2H+2\right]\II_{mn}\partial H \partial H \\
	&=0.
\end{aligned}
\end{equation}

We have thus shown that all the terms in $K_{MN}$ and $R$ vanish and the equations of motion are therefore satisfied by our solution.

\addcontentsline{toc}{section}{References}
\bibliographystyle{JHEP} 
\bibliography{mybib}

\providecommand{\href}[2]{#2}\begingroup\raggedright\begin{thebibliography}{10}

\bibitem{Townsend:1995kk}
P.~Townsend, {\it {The eleven-dimensional supermembrane revisited}},  {\em
  Phys.Lett.} {\bf B350} (1995) 184--187,
  [\href{http://xxx.lanl.gov/abs/hep-th/9501068}{{\tt hep-th/9501068}}].

\bibitem{Witten:1995ex}
E.~Witten, {\it {String theory dynamics in various dimensions}},  {\em
  Nucl.Phys.} {\bf B443} (1995) 85--126,
  [\href{http://xxx.lanl.gov/abs/hep-th/9503124}{{\tt hep-th/9503124}}].

\bibitem{Townsend:1997wg}
P.~Townsend, {\it {M theory from its superalgebra}},
  \href{http://xxx.lanl.gov/abs/hep-th/9712004}{{\tt hep-th/9712004}}.

\bibitem{Hull:2009mi}
C.~Hull and B.~Zwiebach, {\it {Double Field Theory}},  {\em JHEP} {\bf 0909}
  (2009) 099, [\href{http://xxx.lanl.gov/abs/0904.4664}{{\tt
  arXiv:0904.4664}}].

\bibitem{Hull:2009zb}
C.~Hull and B.~Zwiebach, {\it {The Gauge Algebra of Double Field Theory and
  Courant Brackets}},  {\em JHEP} {\bf 0909} (2009) 090,
  [\href{http://xxx.lanl.gov/abs/0908.1792}{{\tt arXiv:0908.1792}}].

\bibitem{Hohm:2010jy}
O.~Hohm, C.~Hull, and B.~Zwiebach, {\it {Background Independent Action for
  Double Field Theory}},  {\em JHEP} {\bf 1007} (2010) 016,
  [\href{http://xxx.lanl.gov/abs/1003.5027}{{\tt arXiv:1003.5027}}].

\bibitem{Hohm:2010pp}
O.~Hohm, C.~Hull, and B.~Zwiebach, {\it {Generalized Metric Formulation of
  Double Field Theory}},  {\em JHEP} {\bf 1008} (2010) 008,
  [\href{http://xxx.lanl.gov/abs/1006.4823}{{\tt arXiv:1006.4823}}].

\bibitem{Jeon:2010rw}
I.~Jeon, K.~Lee, and J.-H. Park, {\it {Differential Geometry with a Projection:
  Application to Double Field Theory}},  {\em JHEP} {\bf 1104} (2011) 014,
  [\href{http://xxx.lanl.gov/abs/1011.1324}{{\tt arXiv:1011.1324}}].

\bibitem{Jeon:2011cn}
I.~Jeon, K.~Lee, and J.-H. Park, {\it {Stringy Differential Geometry, Beyond
  Riemann}},  {\em Phys.Rev.} {\bf D84} (2011) 044022,
  [\href{http://xxx.lanl.gov/abs/1105.6294}{{\tt arXiv:1105.6294}}].

\bibitem{Jeon:2011vx}
I.~Jeon, K.~Lee, and J.-H. Park, {\it {Incorporation of Fermions into Double
  Field Theory}},  {\em JHEP} {\bf 1111} (2011) 025,
  [\href{http://xxx.lanl.gov/abs/1109.2035}{{\tt arXiv:1109.2035}}].

\bibitem{Jeon:2011sq}
I.~Jeon, K.~Lee, and J.-H. Park, {\it {Supersymmetric Double Field Theory:
  Stringy Reformulation of Supergravity}},  {\em Phys.Rev.} {\bf D85} (2012)
  081501, [\href{http://xxx.lanl.gov/abs/1112.0069}{{\tt arXiv:1112.0069}}].

\bibitem{Jeon:2012hp}
I.~Jeon, K.~Lee, J.-H. Park, and Y.~Suh, {\it {Stringy Unification of Type IIA
  and IIB Supergravities under N=2 D=10 Supersymmetric Double Field Theory}},
  \href{http://xxx.lanl.gov/abs/1210.5078}{{\tt arXiv:1210.5078}}.

\bibitem{Aldazabal:2011nj}
G.~Aldazabal, W.~Baron, D.~Marqu\'{e}s, and C.~N\'{u}\~{n}ez, {\it {The
  Effective Action of Double Field Theory}},  {\em JHEP} {\bf 1111} (2011) 052,
  [\href{http://xxx.lanl.gov/abs/1109.0290}{{\tt arXiv:1109.0290}}].

\bibitem{Coimbra:2011nw}
A.~Coimbra, C.~Strickland-Constable, and D.~Waldram, {\it {Supergravity as
  Generalised Geometry I: Type II Theories}},  {\em JHEP} {\bf 1111} (2011)
  091, [\href{http://xxx.lanl.gov/abs/1107.1733}{{\tt arXiv:1107.1733}}].

\bibitem{Aldazabal:2013sca}
G.~Aldazabal, D.~Marqu\'{e}s, and C.~N\'{u}\~{n}ez, {\it {Double Field Theory:
  A Pedagogical Review}},  \href{http://xxx.lanl.gov/abs/1305.1907}{{\tt
  arXiv:1305.1907}}.

\bibitem{Berman:2013eva}
D.~S. Berman and D.~C. Thompson, {\it {Duality Symmetric String and M-Theory}},
   \href{http://xxx.lanl.gov/abs/1306.2643}{{\tt arXiv:1306.2643}}.

\bibitem{Hohm:2013bwa}
O.~Hohm, D.~Lust, and B.~Zwiebach, {\it {The Spacetime of Double Field Theory:
  Review, Remarks, and Outlook}},  {\em Fortsch.Phys.} {\bf 61} (2013)
  926--966, [\href{http://xxx.lanl.gov/abs/1309.2977}{{\tt arXiv:1309.2977}}].

\bibitem{Hohm:2012mf}
O.~Hohm and B.~Zwiebach, {\it {Towards an Invariant Geometry of Double Field
  Theory}},  \href{http://xxx.lanl.gov/abs/1212.1736}{{\tt arXiv:1212.1736}}.

\bibitem{Berman:2014jba}
D.~S. Berman, M.~Cederwall, and M.~J. Perry, {\it {Global aspects of double
  geometry}},  \href{http://xxx.lanl.gov/abs/1401.1311}{{\tt arXiv:1401.1311}}.

\bibitem{Cederwall:2014kxa}
M.~Cederwall, {\it {The geometry behind double geometry}},
  \href{http://xxx.lanl.gov/abs/1402.2513}{{\tt arXiv:1402.2513}}.

\bibitem{Hillmann:2009ci}
C.~Hillmann, {\it {Generalized E(7(7)) coset dynamics and D=11 supergravity}},
  {\em JHEP} {\bf 0903} (2009) 135,
  [\href{http://xxx.lanl.gov/abs/0901.1581}{{\tt arXiv:0901.1581}}].

\bibitem{Hull:2007zu}
C.~Hull, {\it {Generalised Geometry for M-Theory}},  {\em JHEP} {\bf 0707}
  (2007) 079, [\href{http://xxx.lanl.gov/abs/hep-th/0701203}{{\tt
  hep-th/0701203}}].

\bibitem{Pacheco:2008ps}
P.~P. Pacheco and D.~Waldram, {\it {M-theory, exceptional generalised geometry
  and superpotentials}},  {\em JHEP} {\bf 0809} (2008) 123,
  [\href{http://xxx.lanl.gov/abs/0804.1362}{{\tt arXiv:0804.1362}}].

\bibitem{Coimbra:2011ky}
A.~Coimbra, C.~Strickland-Constable, and D.~Waldram, {\it {$E_{d(d)} \times
  \mathbb{R}^+$ generalised geometry, connections and M theory}},  {\em JHEP}
  {\bf 1402} (2014) 054, [\href{http://xxx.lanl.gov/abs/1112.3989}{{\tt
  arXiv:1112.3989}}].

\bibitem{Coimbra:2012af}
A.~Coimbra, C.~Strickland-Constable, and D.~Waldram, {\it {Supergravity as
  Generalised Geometry II: $E_{d(d)}\times \mathbb{R}^+$ and M theory}},  {\em
  JHEP} {\bf 1403} (2014) 019, [\href{http://xxx.lanl.gov/abs/1212.1586}{{\tt
  arXiv:1212.1586}}].

\bibitem{Berman:2010is}
D.~S. Berman and M.~J. Perry, {\it {Generalized Geometry and M theory}},  {\em
  JHEP} {\bf 1106} (2011) 074, [\href{http://xxx.lanl.gov/abs/1008.1763}{{\tt
  arXiv:1008.1763}}].

\bibitem{Berman:2011pe}
D.~S. Berman, H.~Godazgar, and M.~J. Perry, {\it {SO(5,5) duality in M-theory
  and generalized geometry}},  {\em Phys.Lett.} {\bf B700} (2011) 65--67,
  [\href{http://xxx.lanl.gov/abs/1103.5733}{{\tt arXiv:1103.5733}}].

\bibitem{Berman:2011jh}
D.~S. Berman, H.~Godazgar, M.~J. Perry, and P.~West, {\it {Duality Invariant
  Actions and Generalised Geometry}},  {\em JHEP} {\bf 1202} (2012) 108,
  [\href{http://xxx.lanl.gov/abs/1111.0459}{{\tt arXiv:1111.0459}}].

\bibitem{Berman:2011cg}
D.~S. Berman, H.~Godazgar, M.~Godazgar, and M.~J. Perry, {\it {The Local
  symmetries of M-theory and their formulation in generalised geometry}},  {\em
  JHEP} {\bf 1201} (2012) 012, [\href{http://xxx.lanl.gov/abs/1110.3930}{{\tt
  arXiv:1110.3930}}].

\bibitem{Berman:2012vc}
D.~S. Berman, M.~Cederwall, A.~Kleinschmidt, and D.~C. Thompson, {\it {The
  Gauge Structure of Generalised Diffeomorphisms}},  {\em JHEP} {\bf 1301}
  (2013) 064, [\href{http://xxx.lanl.gov/abs/1208.5884}{{\tt
  arXiv:1208.5884}}].

\bibitem{Dabholkar:1990yf}
A.~Dabholkar, G.~W. Gibbons, J.~A. Harvey, and F.~Ruiz~Ruiz, {\it {Superstrings
  and Solitons}},  {\em Nucl.Phys.} {\bf B340} (1990) 33--55.

\bibitem{Adawi:1998ta}
T.~Adawi, M.~Cederwall, U.~Gran, B.~E. Nilsson, and B.~Razaznejad, {\it
  {Goldstone tensor modes}},  {\em JHEP} {\bf 9902} (1999) 001,
  [\href{http://xxx.lanl.gov/abs/hep-th/9811145}{{\tt hep-th/9811145}}].

\bibitem{Tseytlin90}
A.~A. Tseytlin, {\it {Duality Symmetric Formulation of String World Sheet
  Dynamics}},  {\em Phys.Lett.} {\bf B242} (1990) 163--174.

\bibitem{Tseytlin91}
A.~A. Tseytlin, {\it {Duality Symmetric Closed String Theory and Interacting
  Chiral Scalars}},  {\em Nucl.Phys.} {\bf B350} (1991) 395--440.

\bibitem{Siegel93a}
W.~Siegel, {\it {Two Vierbein Formalism for String Inspired Axionic Gravity}},
  {\em Phys.Rev.} {\bf D47} (1993) 5453--5459,
  [\href{http://xxx.lanl.gov/abs/hep-th/9302036}{{\tt hep-th/9302036}}].

\bibitem{Siegel93b}
W.~Siegel, {\it {Superspace Duality in Low-Energy Superstrings}},  {\em
  Phys.Rev.} {\bf D48} (1993) 2826--2837,
  [\href{http://xxx.lanl.gov/abs/hep-th/9305073}{{\tt hep-th/9305073}}].

\bibitem{Duff90a}
M.~Duff, {\it {Duality Rotations in String Theory}},  {\em Nucl.Phys.} {\bf
  B335} (1990) 610.

\bibitem{Duff90b}
M.~Duff and J.~Lu, {\it {Duality Rotations in Membrane Theory}},  {\em
  Nucl.Phys.} {\bf B347} (1990) 394--419.

\bibitem{Aldazabal:2013via}
G.~Aldazabal, M.~Gra\~{n}a, D.~Marqu\'{e}s, and J.~A. Rosabal, {\it {The gauge
  structure of Exceptional Field Theories and the tensor hierarchy}},
  \href{http://xxx.lanl.gov/abs/1312.4549}{{\tt arXiv:1312.4549}}.

\bibitem{Hohm:2013pua}
O.~Hohm and H.~Samtleben, {\it {Exceptional Form of D=11 Supergravity}},  {\em
  Phys.Rev.Lett.} {\bf 111} (2013) 231601,
  [\href{http://xxx.lanl.gov/abs/1308.1673}{{\tt arXiv:1308.1673}}].

\bibitem{Hohm:2013vpa}
O.~Hohm and H.~Samtleben, {\it {Exceptional Field Theory I: $E_{6(6)}$
  covariant Form of M-Theory and Type IIB}},
  \href{http://xxx.lanl.gov/abs/1312.0614}{{\tt arXiv:1312.0614}}.

\bibitem{Hohm:2013uia}
O.~Hohm and H.~Samtleben, {\it {Exceptional Field Theory II: E$_{7(7)}$}},
  \href{http://xxx.lanl.gov/abs/1312.4542}{{\tt arXiv:1312.4542}}.

\bibitem{West:2001as}
P.~C. West, {\it {E(11) and M theory}},  {\em Class.Quant.Grav.} {\bf 18}
  (2001) 4443--4460, [\href{http://xxx.lanl.gov/abs/hep-th/0104081}{{\tt
  hep-th/0104081}}].

\bibitem{Englert:2003zs}
F.~Englert, L.~Houart, A.~Taormina, and P.~C. West, {\it {The Symmetry of M
  theories}},  {\em JHEP} {\bf 0309} (2003) 020,
  [\href{http://xxx.lanl.gov/abs/hep-th/0304206}{{\tt hep-th/0304206}}].

\bibitem{West:2003fc}
P.~C. West, {\it {E(11), SL(32) and central charges}},  {\em Phys.Lett.} {\bf
  B575} (2003) 333--342, [\href{http://xxx.lanl.gov/abs/hep-th/0307098}{{\tt
  hep-th/0307098}}].

\bibitem{Kleinschmidt:2003jf}
A.~Kleinschmidt and P.~C. West, {\it {Representations of G+++ and the role of
  space-time}},  {\em JHEP} {\bf 0402} (2004) 033,
  [\href{http://xxx.lanl.gov/abs/hep-th/0312247}{{\tt hep-th/0312247}}].

\bibitem{West:2004kb}
P.~C. West, {\it {E(11) origin of brane charges and U-duality multiplets}},
  {\em JHEP} {\bf 0408} (2004) 052,
  [\href{http://xxx.lanl.gov/abs/hep-th/0406150}{{\tt hep-th/0406150}}].

\bibitem{West:2012qm}
P.~West, {\it {Generalised BPS conditions}},  {\em Mod.Phys.Lett.} {\bf A27}
  (2012) 1250202, [\href{http://xxx.lanl.gov/abs/1208.3397}{{\tt
  arXiv:1208.3397}}].

\bibitem{Berman:2007vi}
D.~S. Berman and N.~B. Copland, {\it {The String partition function in Hull's
  doubled formalism}},  {\em Phys.Lett.} {\bf B649} (2007) 325--333,
  [\href{http://xxx.lanl.gov/abs/hep-th/0701080}{{\tt hep-th/0701080}}].

\bibitem{Berman:2007xn}
D.~S. Berman, N.~B. Copland, and D.~C. Thompson, {\it {Background Field
  Equations for the Duality Symmetric String}},  {\em Nucl.Phys.} {\bf B791}
  (2008) 175--191, [\href{http://xxx.lanl.gov/abs/0708.2267}{{\tt
  arXiv:0708.2267}}].

\bibitem{Berman:2007yf}
D.~S. Berman and D.~C. Thompson, {\it {Duality Symmetric Strings, Dilatons and
  O(d,d) Effective Actions}},  {\em Phys.Lett.} {\bf B662} (2008) 279--284,
  [\href{http://xxx.lanl.gov/abs/0712.1121}{{\tt arXiv:0712.1121}}].

\bibitem{Hohm:2013jaa}
O.~Hohm, W.~Siegel, and B.~Zwiebach, {\it {Doubled $\alpha'$-geometry}},  {\em
  JHEP} {\bf 1402} (2014) 065, [\href{http://xxx.lanl.gov/abs/1306.2970}{{\tt
  arXiv:1306.2970}}].

\bibitem{Betz:2014aia}
A.~Betz, R.~Blumenhagen, D.~Lust, and F.~Rennecke, {\it {A Note on the CFT
  Origin of the Strong Constraint of DFT}},
  \href{http://xxx.lanl.gov/abs/1402.1686}{{\tt arXiv:1402.1686}}.

\bibitem{Berman:2013uda}
D.~S. Berman, C.~D.~A. Blair, E.~Malek, and M.~J. Perry, {\it {The $O_{D,D}$
  Geometry of String Theory}},  \href{http://xxx.lanl.gov/abs/1303.6727}{{\tt
  arXiv:1303.6727}}.

\bibitem{Blair:2013noa}
C.~D.~A. Blair, E.~Malek, and A.~J. Routh, {\it {An O(D,D) Invariant
  Hamiltonian Action for the Superstring}},
  \href{http://xxx.lanl.gov/abs/1308.4829}{{\tt arXiv:1308.4829}}.

\bibitem{Blair:2013gqa}
C.~D.~A. Blair, E.~Malek, and J.-H. Park, {\it {M-theory and Type IIB from a
  Duality Manifest Action}},  {\em JHEP} {\bf 1401} (2014) 172,
  [\href{http://xxx.lanl.gov/abs/1311.5109}{{\tt arXiv:1311.5109}}].

\bibitem{Lee:2014mla}
K.~Lee, C.~Strickland-Constable, and D.~Waldram, {\it {Spheres, generalised
  parallelisability and consistent truncations}},
  \href{http://xxx.lanl.gov/abs/1401.3360}{{\tt arXiv:1401.3360}}.

\bibitem{Lee:2013hma}
K.~Lee and J.-H. Park, {\it {Covariant action for a string in doubled yet
  gauged spacetime}},  {\em Nucl.Phys.} {\bf B880} (2014) 134--154,
  [\href{http://xxx.lanl.gov/abs/1307.8377}{{\tt arXiv:1307.8377}}].

\bibitem{Strickland-Constable:2013xta}
C.~Strickland-Constable, {\it {Subsectors, Dynkin Diagrams and New Generalised
  Geometries}},  \href{http://xxx.lanl.gov/abs/1310.4196}{{\tt
  arXiv:1310.4196}}.

\bibitem{Park:2014una}
J.-H. Park and Y.~Suh, {\it {U-gravity : ${\mathbf{SL}(N)}$}},
  \href{http://xxx.lanl.gov/abs/1402.5027}{{\tt arXiv:1402.5027}}.

\bibitem{Cederwall:2013naa}
M.~Cederwall, J.~Edlund, and A.~Karlsson, {\it {Exceptional geometry and tensor
  fields}},  {\em JHEP} {\bf 1307} (2013) 028,
  [\href{http://xxx.lanl.gov/abs/1302.6736}{{\tt arXiv:1302.6736}}].

\bibitem{Ortin04}
T.~Ort\'{i}n, {\em {Gravity and Strings}}.
\newblock Cambridge University Press, 2004.

\bibitem{Berman:2011kg}
D.~S. Berman, E.~T. Musaev, and M.~J. Perry, {\it {Boundary Terms in
  Generalized Geometry and doubled field theory}},  {\em Phys.Lett.} {\bf B706}
  (2011) 228--231, [\href{http://xxx.lanl.gov/abs/1110.3097}{{\tt
  arXiv:1110.3097}}].

\bibitem{Berman:2012uy}
D.~S. Berman, E.~T. Musaev, and D.~C. Thompson, {\it {Duality Invariant
  M-theory: Gauged supergravities and Scherk-Schwarz reductions}},  {\em JHEP}
  {\bf 1210} (2012) 174, [\href{http://xxx.lanl.gov/abs/1208.0020}{{\tt
  arXiv:1208.0020}}].

\bibitem{Berman:2013cli}
D.~S. Berman and K.~Lee, {\it {Supersymmetry for Gauged Double Field Theory and
  Generalised Scherk-Schwarz Reductions}},  {\em Nucl.Phys.} {\bf B881} (2014)
  369--390, [\href{http://xxx.lanl.gov/abs/1305.2747}{{\tt arXiv:1305.2747}}].

\bibitem{Grana:2012rr}
M.~Gra\~{n}a and D.~Marqu\'{e}s, {\it {Gauged Double Field Theory}},  {\em
  JHEP} {\bf 1204} (2012) 020, [\href{http://xxx.lanl.gov/abs/1201.2924}{{\tt
  arXiv:1201.2924}}].

\bibitem{Geissbuhler:2011mx}
D.~Geissbuhler, {\it {Double Field Theory and N=4 Gauged Supergravity}},  {\em
  JHEP} {\bf 1111} (2011) 116, [\href{http://xxx.lanl.gov/abs/1109.4280}{{\tt
  arXiv:1109.4280}}].

\bibitem{Aldazabal:2013mya}
G.~Aldazabal, M.~Gra\~{n}a, D.~Marqu\'{e}s, and J.~Rosabal, {\it {Extended
  Geometry and Gauged Maximal Supergravity}},
  \href{http://xxx.lanl.gov/abs/1302.5419}{{\tt arXiv:1302.5419}}.

\bibitem{Aichelburg:1970dh}
P.~Aichelburg and R.~Sexl, {\it {On the Gravitational field of a massless
  particle}},  {\em Gen.Rel.Grav.} {\bf 2} (1971) 303--312.

\bibitem{Kaplan:1995cp}
D.~M. Kaplan and J.~Michelson, {\it {Zero modes for the D = 11 membrane and
  five-brane}},  {\em Phys.Rev.} {\bf D53} (1996) 3474--3476,
  [\href{http://xxx.lanl.gov/abs/hep-th/9510053}{{\tt hep-th/9510053}}].

\bibitem{Malek:2012pw}
E.~Malek, {\it {U-duality in three and four dimensions}},
  \href{http://xxx.lanl.gov/abs/1205.6403}{{\tt arXiv:1205.6403}}.

\bibitem{ArkaniHamed:2012nw}
N.~Arkani-Hamed, J.~L. Bourjaily, F.~Cachazo, A.~B. Goncharov, A.~Postnikov,
  et~al., {\it {Scattering Amplitudes and the Positive Grassmannian}},
  \href{http://xxx.lanl.gov/abs/1212.5605}{{\tt arXiv:1212.5605}}.

\bibitem{Jensen:2011jna}
S.~Jensen, {\it {The KK-Monopole/NS5-Brane in Doubled Geometry}},  {\em JHEP}
  {\bf 1107} (2011) 088, [\href{http://xxx.lanl.gov/abs/1106.1174}{{\tt
  arXiv:1106.1174}}].

\end{thebibliography}\endgroup

\end{document}